\documentclass[apj]{emulateapj}
\usepackage{graphicx}
\usepackage[flushleft]{threeparttable}
\usepackage[usenames,dvipsnames,svgnames,table]{xcolor}
\usepackage{hyperref}
\definecolor{darkblue}{rgb}{0.0,0.0,0.3}
\hypersetup{colorlinks,breaklinks,
            linkcolor=darkblue,urlcolor=darkblue,
            anchorcolor=darkblue,citecolor=darkblue}
\usepackage{amsmath,amssymb}
\usepackage{amsmath}

\usepackage{natbib}
\begin{document}

\def\etal{et al.\ \rm}
\def\ba{\begin{eqnarray}}
\def\ea{\end{eqnarray}}
\def\etal{et al.\ \rm}

\title{Spreading Layers in Accreting Objects: Role of Acoustic Waves for Angular Momentum Transport, Mixing and Thermodynamics}

\author{Alexander A. Philippov\altaffilmark{1,3}, Roman R. Rafikov\altaffilmark{1,2} \&
James M. Stone\altaffilmark{1}}

\altaffiltext{1}{Department of Astrophysical Sciences, 
Princeton University, Ivy Lane, Princeton, NJ 08540}
\altaffiltext{2}{Institute for Advanced Study, Einstein Drive, Princeton, NJ 08540}
\altaffiltext{3}{sashaph@princeton.edu}


\begin{abstract}
Disk accretion at high rate onto a white dwarf or a neutron star has been suggested to result in the formation of a spreading layer (SL) --- a belt-like structure on the object's surface, in which the accreted matter steadily spreads in the poleward (meridional) direction while spinning down. To assess its basic characteristics we perform two-dimensional hydrodynamic simulations of supersonic SLs in the relevant morphology with a simple prescription for cooling. We demonstrate that supersonic shear naturally present at the base of the SL inevitably drives sonic instability that gives rise to large scale acoustic modes governing the evolution of the SL. These modes dominate the transport of momentum and energy, which is intrinsically global and cannot be characterized via some form of local effective viscosity (e.g. $\alpha$-viscosity). The global nature of the wave-driven transport {should} have  important implications for triggering Type I X-ray bursts in low mass X-ray binaries. The nonlinear evolution of waves into a system of shocks drives effective re-arrangement ({sensitively depending on thermodynamical properties of the flow}) and deceleration of the SL, which ultimately becomes transonic and susceptible to regular Kelvin-Helmholtz instability. We interpret this evolution in terms of the global structure of the SL and suggest that mixing of the SL material with the underlying stellar fluid should become effective only at intermediate latitudes {on the accreting object's surface}, where the flow has decelerated appreciably. In the near-equatorial regions the transport is dominated by acoustic waves and mixing is less efficient. We speculate that this latitudinal non-uniformity of mixing in accreting white dwarfs may be linked to the observed bipolar morphology of classical novae ejecta.
\end{abstract}



\section{Introduction.}  
\label{sect:intro}

Disk accretion onto astrophysical objects possessing a physical surface (as opposed to accretion onto black holes) inevitably involves passage of matter from the inner parts of the disk onto the surface of the accreting object. The central object could be a white dwarf (WD) in a cataclysmic variable (CV), an accreting neutron star (NS) in a low-mass X-ray binary (LMXB), a young star accreting from a protoplanetary disk, or a growing giant planet accreting from a circumplanetary disk. 

When the magnetic field of the central object is high enough, the inner disk is disrupted, and a cavity forms in its center. In this case the flow proceeds along the magnetic field lines \citep{Ghosh}, which channel the gas onto the stellar\footnote{We will use the term "star" very broadly in this work, implying any accreting astrophysical object with a surface.} surface. 

However, in many astrophysically relevant situations the field is too weak or the mass accretion rate $\dot M$ is too high for the flow to be disrupted by the magnetic stresses. Such a situation is typical for weakly magnetized neutron stars in LMXBs and white dwarfs in CVs, as well as for young stars undergoing FU Orioni outbursts. In this case the accretion disk extends all the way towards the stellar surface and a boundary layer (BL) inevitably forms where the two meet. By definition, the BL is the inner part of the accretion disc where the flow is decelerated from the Keplerian speed in the disk to the stellar rotational velocity. This azimuthal velocity drop occurs mainly in the radial direction in the BL, which motivated the development of the early one-dimensional (1D), vertically integrated models of the BLs \citep{LyndenBell74, Pringle77, Tylenda81, Regev83, NarPop, PophamPMS, PophamWD}. 

A different morphology of the transition between the disk and the stellar surface was suggested by \citet{InogamovSunyaev} in the context of disk accretion onto the weakly magnetized neutron stars. They hypothesized that when $\dot M$ is high enough and the density inside the star increases rapidly, the accreted material does not attain stellar rotation rate immediately but instead spreads meridionally as it gradually loses its angular momentum. This leads to the formation of the so-called {\it spreading layer} (SL) on the stellar surface, which can extend rather far in latitude from the disk plane. In this picture SL has small radial thickness and its characteristics vary mainly in the meridional direction, motivating 1D, radially-integrated description \citep{InogamovSunyaev,IS2}. Later the concept of the SL was extended to the case of accretion onto the white dwarfs \citep{Piro}.

Since the rotational velocity of the gas passing through the BL (and SL) changes with
radius (meridional distance), some mechanism of angular momentum transport must operate
in the layer, ultimately resulting also in mass transport. Magnetorotational
instability (MRI, \citet{Velikhov, Chandra}) has been successfully invoked to explain angular momentum transport in the bulk of the disk \citep{MRI}. {However, it} does not operate in the BL region since the necessary condition for it to work is $d\Omega^2/dr<0$ ($\Omega$ is the angular frequency of the fluid), which is obviously not fulfilled in the BL. Thus, angular momentum transport in the BL must be effected by the means other than MRI. In the absence of proper understanding of the momentum transport in the BL (and SL), numerical models used different prescriptions for the local artificial viscosity \citep{Kley89, Balsara05, Balsara09, Hertfelder13}. Despite successfully producing solutions which can be compared with observations \citep{Suleimanov14}, these models do not explain the physical nature of the momentum transport in the BLs.  

Recently \citet{Belyaev1} discovered a completely different mechanism of angular
momentum transport in the BL. They found that supersonic shear naturally present at the interface between the disk and the star excites global non-axisymmetric acoustic modes. These modes, with pressure as their restoring force, are excited through the sonic instability, which is a type of shear instability that occurs in supersonic, highly compressible
flows \citep{Glatzel,Belyaev0}. This instability is distinct from the regular Kelvin-Helmholtz (hereafter KH) instability, which is known to have difficulty operating in supersonic shear flows \citep{Miles1958}. \citet{Belyaev1} demonstrated that natural evolution of these acoustic modes into weak  shocks results in their dissipation and non-local transport of energy and angular momentum in the BL region. This eventually gives rise to mass transport resulting in accretion onto the star. The global nature of this transport precludes its description via some local, effective $\alpha$-prescription for anomalous viscosity. In \citet{Belyaev2} and \citet{Belyaev3} it was also shown that taking into account 3D effects and magnetic fields does not change the general picture of the sonic wave-mediated transport in the astrophysical BLs. 

The main goal of our present work is to demonstrate that acoustic modes provide a natural mechanism of the momentum and energy transport not only in the BLs but also in {the} SLs \citep{InogamovSunyaev}. It is not immediately obvious that this should be the case in the geometric setup of the SL and we explore this issue with high-resolution hydrodynamical simulations (we leave study of the magnetic effects for the future).

Another goal of this work is to probe the effect of the non-trivial disk {\it thermodynamics} on the operation of the acoustic wave-driven transport. \citet{Belyaev1,Belyaev2,Belyaev3} explored this instability using only isothermal equation of state (EOS). In this paper we relax this assumption and study various prescriptions for gas thermodynamics: isothermal EOS, adiabatic EOS, and the possibility of gas cooling at a prescribed constant rate. This allows us to study the {\it back-reaction} of the energy transport on the temperature distribution and sonic wave propagation in the SL. Similar investigation has been performed recently in the BL context by \citet{Kley}, who found that sonic instability and wave-mediated transport persist even with non-trivial thermodynamics.

In addition, we also study the efficiency of vertical {\it mixing} inside the SL by tracing passive scalar in the simulations. Previously, turbulent mixing between the underlying layers of the WD and freshly accreted material, which have different compositions, was explored \citep{Truran} to explain the observed enrichment of novae ejecta in heavy elements \citep{Gehrz}, as well as to assess the possible role of the single-degenerate scenario for triggering Type Ia supernovae \citep{sn1a}. A number of instabilities that could generate turbulence and lead to efficient mixing have been explored in the subsonic regime \citep{Alexakis01,Alexakis04,Casanova10,Casanova11}. However, global acoustic modes naturally emerging in the supersonic shear flows generally do not lead to excitation of large scale turbulent motions, at least in the BL setup \citep{Belyaev1,Belyaev2}. This motivates us to explore the role of mixing in the physics of the supersonic SLs in more details. 

The paper is organized as follows. In Section \ref{sect:setup} we describe our geometric, physical and numerical setup, as well as the boundary and initial conditions. After briefly discussing the linear development of the sonic instability in the SL setup in Section \ref{sect:WDaccretion}, we provide detailed description of our results in the non-linear phase for isothermal (Section \ref{sect:WDisotherm}), no cooling (Section \ref{sec:adiabatic}), and some cooling (Section \ref{sect:with_cooling}) simulations. In doing so we focus on detailed characterization of the properties of the wave-driven transport, mixing, generation of vorticity, etc. We discuss our results in Section \ref{sect:discussion}, focusing on the role of mixing (\S \ref{sect:mix}), energy transport (\S \ref{sect:en_transport}), global SL characteristics (\S \ref{sect:global_outlook}), and comparison with previous work (\S \ref{sect:compare}). Applications of our results to Type I X-ray bursts and bipolar nova explosions are discussed in \S \ref{sect:en_transport} and \S \ref{sect:bipolar}, correspondingly. Our main conclusions are summarized in Section \ref{sect:conclusions}.


\section{Problem setup.}  
\label{sect:setup}


We start by providing a detailed description of the SL geometry and our numerical setup used to explore the SL properties.


\subsection{Geometry of the problem.}  
\label{sect:geometry}


\begin{figure}
\epsscale{1.}
\plotone{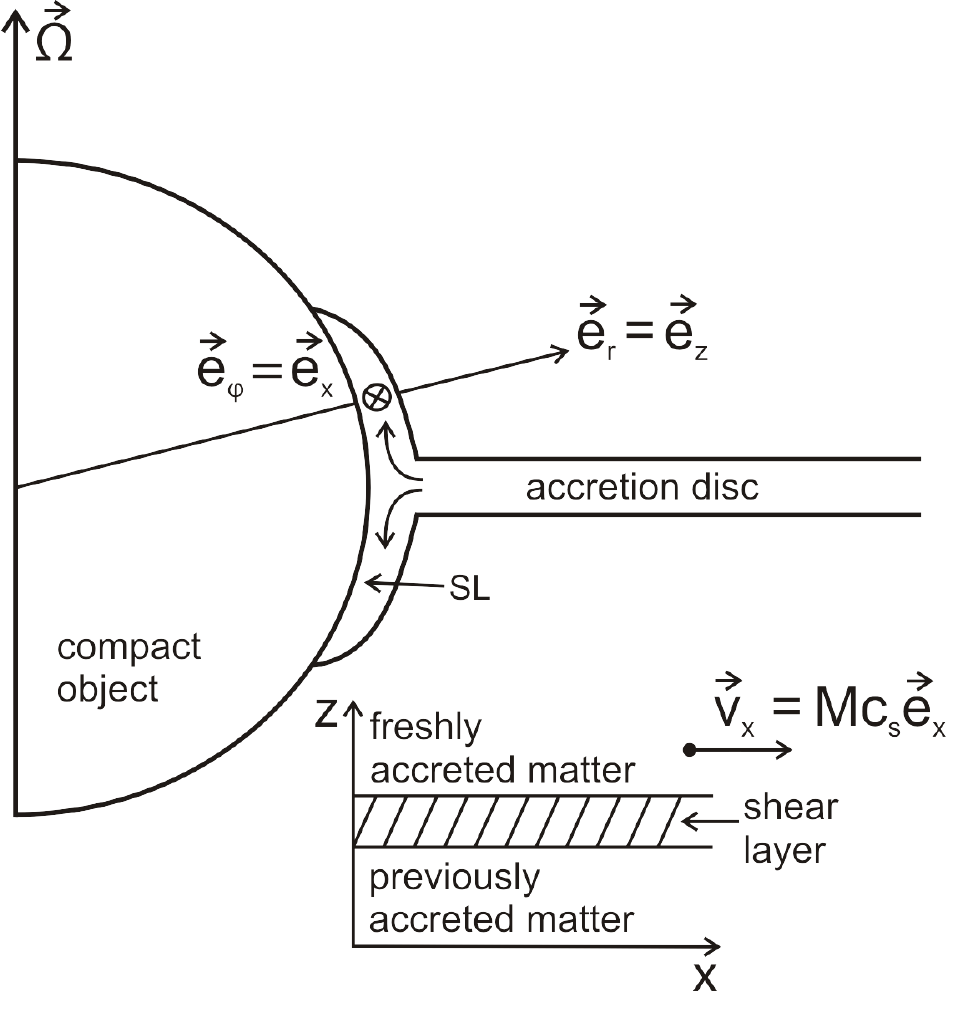}
\caption{Description of the geometry and coordinate system used in this work. Our 2D simulation domain is aligned perpendicular to the meridional plane of the spherical coordinate system, with ${\bf e}_z$ and ${\bf e}_x$ axes aligned with the local ${\bf e}_r$ and ${\bf e}_\phi$, correspondingly. A more detailed view of the box {in the orthogonal plane} with its main constituent parts is shown in the bottom of the Figure.}
\label{figcoord}
\end{figure}

Fully global investigation of the SL can be carried out most naturally in standard spherical $(r,\theta,\phi)$ coordinates. However, given the exploratory nature of our pilot study of the sonic instability in the SL and significant computational demands, in this work we adopt a local approximation and consider only a small section of the SL.

Moreover, we also reduce the dimensionality of the problem from 3D to 2D by focusing on fluid  dynamics only at fixed meridional angle $\theta$, in the plane normal to the direction of the poleward spreading, see Figure \ref{figcoord}. This is a serious simplification, which does not allow us to capture the effect of dissipation on the meridional spreading of accreted gas \citep{InogamovSunyaev}. As a result, mass and momentum transport in $\theta$ direction that should naturally arise in the 3D setup are neglected here. However, this geometry still allows us to capture the main features of the wave dynamics in the SL and speeds up our calculations. Also, including third (meridional) dimension in our simulations would require a set of the boundary conditions in this direction, and these are not trivial to formulate in the local approximation. 

Final level of simplification adopted in this work reduces $r$-$\phi$ coordinates of the local patch of the SL at fixed $\theta$ to the Cartesian frame $z$-$x$, where $z$-axis runs parallel to ${\bf e}_r$ and $x$-axis is locally aligned with ${\bf e}_\phi$, see Figure \ref{figcoord}. In other words, in our coordinate notation $z$ is the height above the stellar surface (along the radial direction), and $x$ is the coordinate parallel to the surface and azimuthal velocity of the fluid.

Going from curvilinear to Cartesian coordinates results in our neglect of the effects of curvature and Coriolis force. However, \citet{Belyaev0} have previously shown that these effects do not change the physics of the sonic instability and can be neglected in the local approximation. 

We note that because of these simplifications the SL does not achieve a steady state in our model. This is because dissipation in our simulation box should in practice drive the meridional motion of accreted gas towards the poles. In a 3D SL this gas gets replaced by the lower-latitude fluid carrying higher angular momentum, which enforces (quasi-)steady state in the system. By explicitly neglecting the meridional flow for the reasons outlined above and not imposing external forcing in our simulation box we lose the ability to properly describe the steady state SL {\it at a given latutude}. 

Nevertheless, since the evolution of the SL is slow {(compared to the dynamical timescale)} we are still able to understand the development and operation of the sonic instability in this work (\S \ref{sect:WDaccretion}). Also, in \S \ref{sect:global_outlook} we provide a qualitative interpretation of the SL evolution seen in our simulations in terms of the {\it global} (meridionally-dependent) properties of the SL. We leave the detailed exploration of the 3D structure of the SL to future work.


\subsection{Numerical scheme.}  
\label{sect:numerics}


Our simulations employ the grid-based Godunov code Athena \citep{StoneAthena}
to solve the equations of fluid dynamics in Cartesian geometry, including
energy transport equation. These equations are
\begin{eqnarray}
\frac{\partial \rho}{\partial t} + \boldsymbol{\nabla}  \cdot (\rho
\boldsymbol{v}) = 0,\\
\frac{\partial (\rho \boldsymbol{v})}{\partial t} +
\boldsymbol{\nabla} \cdot (\rho
\boldsymbol{v}\boldsymbol{v}) + \boldsymbol{\nabla}P + \rho \boldsymbol{\nabla}\Phi = 0,\\
\frac{\partial E}{\partial t} + \boldsymbol{\nabla} \cdot ((E+P)
\boldsymbol{v}) = -\rho \boldsymbol{v} \cdot \boldsymbol{\nabla}\Phi -
\rho \Lambda.
\label{energy}
\end{eqnarray}
Here $\boldsymbol{v}$, $\rho$, $P$ are the two dimensional velocity, density, and pressure, $\Phi$ is the gravitational potential, $E =P/(\gamma-1) + \rho v^2/2$ is the full energy of the flow (we use adiabatic index $\gamma =5/3$ throughout this work, except for isothermal runs) and $\Lambda$ is the specific cooling rate (see \S \ref{sect:cooling}). In addition, mixing between the freshly accreted and stellar material is explored by following the evolution of passive scalar $S$ advected with the flow.

We consider the gravitational acceleration $g$ to be constant throughout the simulation
domain, which corresponds to a potential
\begin{equation}
\Phi = g z.
\end{equation}
In the absence of stellar rotation $g = GM_*/R^2_*$, but we leave $g$ {as} a free parameter in general, to allow a possibility of including the centrifugal acceleration due to rapidly spinning central object. 

In the case of the NS, radiation pressure plays a major
role in the flow dynamics \citep{InogamovSunyaev}. In this paper we do not include the radiation pressure into consideration leaving {study of its role} for future work.

The main goal of this work is to {\it identify} the mechanism of the angular momentum transport in the SL geometry. For this reason, we do not include any explicit viscosity in our simulations. It is also important for our work that Athena exhibits very low levels of numerical viscosity. This makes it very well suited for studying nonlinear evolution of the sonic modes, as shown in \citet{Dong, Belyaev1, Belyaev2}. High levels of numerical diffusion may be a reason why sonic instability has not been identified in earlier low resolution studies of the BLs \citep{Armitage, Steinacker}.


\subsection{Treatment of thermodynamics.}  
\label{sect:cooling}


In this paper we pay special attention to the effects of {\it thermodynamics} on the wave excitation, propagation and back-reaction on the global properties of the flow. Just to reiterate, here we consider $P$ to be the gas pressure, neglecting contribution from the radiation pressure which can be important in the NS accretion.

Previous studies of sonic instability \citep{Belyaev1} focused primarily on purely dynamical effects and employed a simple isothermal equation of state (EOS).
In this work we go beyond that approach and consider three different thermodynamical setups to get an idea of how thermodynamics affects the existence and operation of acoustic modes.

To provide a baseline, our first setup employs pure isothermal EOS used in \citet{Belyaev1}, so that $P = \rho c^2_s$, with $c_s$ being the sound speed, which is constant across the simulation domain. In this case the energy equation (\ref{energy}) is not used .

Second, we consider a {\it no cooling} setup, in which the full energy in the simulation box $\int (E + \rho \Phi) \rm {d} V$ is conserved. In this case the thermal energy of the fluid is no longer constant, but increases due to shock dissipation. 

Finally, we explore the effect of gas cooling by implementing a simple cooling function
\begin{equation}
\Lambda = - \frac{T- T_0(z)}{\tau_{cool}}.
\label{eq:cool}
\end{equation}
Here $T$ is the gas temperature, $T_0(z)$ is the initial temperature profile, and $\tau_{cool}$ is the cooling time that
we take constant within our domain (i.e. independent of $\rho$ or $T$) in this work. The effect of this simple cooling prescription is to damp any deviations of fluid temperature from the prescribed profile $T_0(z)$ on a characteristic time $\tau_{cool}$.

Cooling prescription (\ref{eq:cool}) can be thought of as corresponding to the optically thin regime, with constant cooling time $\tau_{cool}$. In practice, both in the NS and in the WD cases the flow in the SL is expected to be optically thick \citep{InogamovSunyaev, Piro} at high values of $\dot M$. Accretion onto the WD at low $\dot M$ may result in the optically thin boundary layer \citep{NarPop} but $\tau_{cool}$ would still be a function of $\rho$ or $T$. Thus, our model of $\Lambda$ may apply to realistic SLs only qualitatively. Nevertheless, even with this rather simple cooling prescription we can still obtain a good initial guess on how the non-trivial gas thermodynamics could change the behavior of the sonic instability in the SL. 


\subsection{Numerical grid and units}
\label{sect:units}


Our simulation box has dimensions of $14H \times 20H$ in units of the initial density scale height $H\equiv c_s^2/g$ in the $x$ and $z$ directions, respectively. We use a grid with $3500 \times 4000$ cells in $x$ and $z$. We find that using lower numerical resolution underestimates the growth rate of the sonic instability, but the nonlinear behavior does not depend on the resolution (see also \citet{Belyaev0}). For all our simulations we checked that the time step is short enough for the cooling time to be well resolved.

The distances are scaled by the initial isothermal scale height $H$, and the density by the initial density at the base of the shear layer $\rho_0=1$. Initially the velocity of the SL is $v_x=1$, and the sound speed $c_s=1/M$, where $M$ is the Mach number of the flow above the shear layer. Time is measured in units of scale height crossing time $H/c_s=c_s/g$.  

All our simulations were run at Mach number $M=5$. Previously, \citet{Belyaev0,Belyaev1} found that variation of $M$ does not affect the qualitative picture of the wave-driven transport, as long as $M\gtrsim 1$. Other parameters of our simulations are summarized in Table \ref{table1:sim}. 


\subsection{Initial conditions.}  
\label{eq:IC}


\begin{figure}
\epsscale{1.35}
\plotone{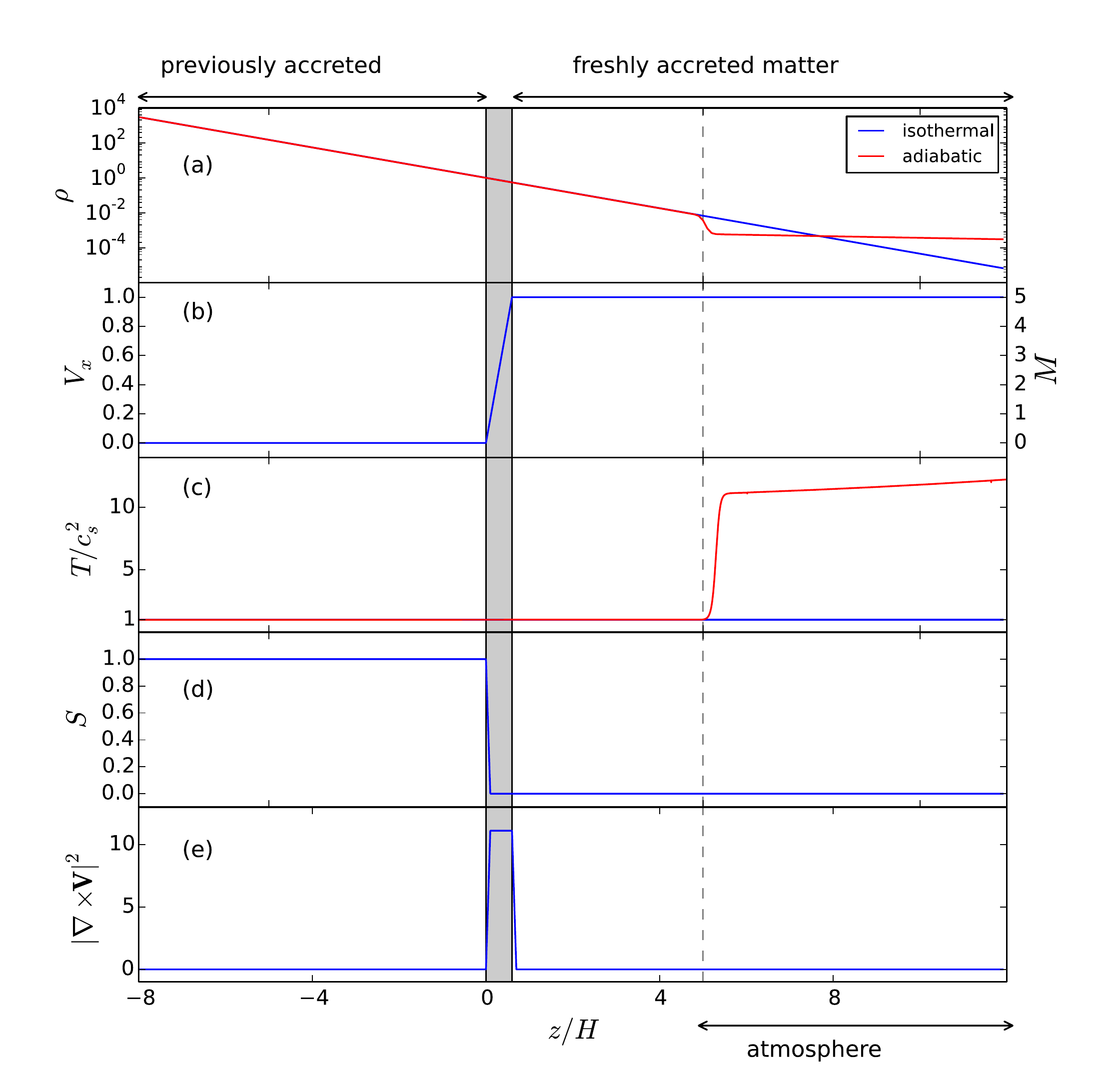}
\caption{Initial configuration of fluid variables in isothermal and non-isothermal (i.e. adiabatic and fixed cooling time) setups (blue and red curves, correspondingly), as a function of (locally) radial coordinate $z$. (a): Density distribution. (b): Profile of the transverse velocity, which is the same for both setups. (c): Temperature distribution, normalized to the {initial} sound speed in the BL region. (d): Distribution of passive scalar. (e): Vorticity distribution. Vertical grey band shows the initial extent of the shear layer.}
\label{fig0}
\end{figure}

The simulation domain in our typical run consists of four (three in the isothermal case) main components located at different altitudes, see Figure \ref{fig0} for illustration and initial profiles of different fluid variables as a function of the vertical coordinate. 

First, at the bottom of the domain we always have a dense layer of {\it previously accreted} fluid, which is initially at rest. We consider this layer to be a part of the star.

Second, immediately above, there is a thin layer in which horizontal speed rapidly rises from zero to a supersonic value $v_x=Mc_s$, matching the velocity of the fluid above. We call this part of the SL a {\it shear layer}.

Third, {\it freshly accreted} fluid moving initially with velocity $v_x=Mc_s$ lies on top of the shear layer. It extends all the way to the top boundary of the simulation box.

Fourth, in non-isothermal runs the upper part of this layer of freshly accreted gas has different thermodynamic properties. We call this region an {\it atmosphere} and discuss it in \S \ref{eq:BC}.

The initial horizontal (azimuthal in the original spherical geometry) velocity profile is thus given by a piece-wise linear approximation:
\ba
    V_x(z)= 
\begin{cases}
    0,& \text{if } z \leq 0\\
    Mc_s(z/h_s),& \text{if } 0\leq z \leq h_s\\
    Mc_s,& \text{otherwise}
\end{cases}
\ea
where $M>1$ is a Mach number of the supersonic flow and $h_s = 0.6~H$ is the initial shear layer width (see also Figure \ref{fig0}b). This velocity profile results in the initial vorticity profile shown in Figure \ref{fig0}e, with non-zero (and constant) vorticity only inside the shear layer.

In real astrophysical situations the initial shear layer width may be set by the microscopic viscosity, which is very low. In this case $h_s\to 0$ and cannot be resolved numerically. For that reason, in our simulations we try a couple of different finite values of $h_s$ to demonstrate that the final, saturated state of the SL does not depend on its exact value, since the shear layer expands in the nonlinear phase of the sonic instability. This independence upon $h_s$ then allows us to expect the saturated state to remain the same in the limit of initial $h_s\to 0$. It is only important to resolve $h_s$ with sufficient number of grid points, to properly capture the physics of the sonic instability \citep{Belyaev0}. Vertical fluid velocity is zero everywhere in the box at the beginning of simulations.

Initial temperature is set to a constant level $T_0$ in isothermal simulations, while in non-isothermal runs it jumps by a factor of $10$ in the atmosphere. 

The initial density profile is specified through the hydrostatic equilibrium condition. For isothermal EOS this results in the exponential density profile
\begin{equation}
\rho = \rho_0 \exp{(-g z/c^2_s)} = \rho_0 \exp{(-z/H)},
\end{equation}
throughout the whole vertical extent of the box. In our non-isothermal runs we adopt the same initial density profile, which implies non-trivial vertical distribution of entropy. Moreover, in this runs density also has a different structure in the atmosphere, see Figure \ref{fig0}.

The initial profile of the passive scalar $S$ is shown on Figure \ref{fig0}d. Penetration of scalar above $z=0$ signals presence of non-trivial vertical mixing.

\begin{table}
\begin{threeparttable}
\caption{Parameters for the simulations}
\begin{tabular}{cccc}
    \hline \hline
    z-range & EOS & BC-z & $\tau_{cool}$ \\ \hline
    (-8;12) & $\gamma=1$ (isothermal) & (do-nothing; reflecting) & 0  \\ 
    (-8;12) & $\gamma=5/3$ (no cooling) & (do-nothing; atmosphere) & $\infty$ \\
    (-8;12) & $\gamma=5/3$  & (do-nothing; atmosphere) & 0.005 \\
    (-8;12) & $\gamma=5/3$  & (do-nothing; atmosphere) & 0.5 \\
    (-8;12) & $\gamma=5/3$  & (do-nothing; atmosphere) & 15. \\
    \hline
\end{tabular}
\begin{tablenotes}
\item {\bf Notes}. z-range: box size in z direction in units of H; EOS: equation of state; BC-z: lower and upper boundary condition in z direction; $\tau_{cool}$: cooling time in units of $H/c_s=c_s/g$. 
\end{tablenotes}
\label{table1:sim}
\end{threeparttable}
\end{table}


\subsection{Boundary conditions.} 
\label{eq:BC}


All our simulations employ standard periodic boundary conditions (BCs) in the $x$-direction. Our BCs in the vertical direction are more complicated and depend on the nature of the accreting object and gas physics, as explained next.

There are two limiting physical setups for the SL we are going to consider. The first one may be relevant for neutron stars, in which case the lower $z$-boundary of the box might be approximated as a solid surface, along which the fluid can move without any friction \citep{InogamovSunyaev}. In order to consider this setup in numerical simulations we impose a {\it reflective} boundary condition at the lower boundary of the simulation domain, above which a non-rotating fluid layer (underlying the shear layer) rests. We refer to this setup as {\it NS BC}. Note that the goal of this approximation is simply to test the effect of a particular BC on the flow. For example, we are not trying to replicate the thermodynamical properties of the SL on the NS surface, which is dominated by the radiation pressure \citep{InogamovSunyaev} not accounted for in our work.

The second setup mimics accretion onto a white dwarf. Here freshly accreted material moves on top of the previously accreted fluid layer of the star, which extends to great depth. In this case we adopt a setup from \citet{Belyaev2} with exponential density profile extended over several scale heights into the star. At the lower boundary of the domain we use the so-called do-nothing BC, which simply means that the fluid variables on this boundary retain their initial values for the duration of the simulation. We refer to this setup as {\it WD BC}.

As for the upper $z$-boundary, we found that it is difficult to arrange a "good" outflow boundary condition because of the mass loss from the simulation box: in our non-isothermal runs gas tends to heat up due to shock dissipation and expand in the vertical direction, leaving the box. To prevent this in our non-isothermal calculations we set up an extended layer of hot gas (referred to hereafter as {\it hot atmosphere}) on top of the dense layer of freshly accreted material, starting at $z_a=5~H$ as shown in Figure \ref{fig0}. The initial temperature of the atmosphere is $\sim 10$ times higher than in the underlying dense layer. As a result, the density in the atmosphere is low (see Figure \ref{fig0}a) suppressing penetration of the sound waves into the atmosphere. Also, density of the atmosphere is almost constant because of its large scale height, so that the waves propagate in it without significant amplification. 

On top of the hot atmosphere we place a reflecting BC. Waves that manage to penetrate the atmosphere and reflect off from the upper side of the simulation box get sufficiently damped on their way back into the region of higher density around the shear layer. As a result, they do not affect the development of sonic instability there. 

For a purely isothermal simulation we can not arrange such an atmosphere since the temperature profile should be constant in these runs. Instead, we just set up a reflecting BC at the upper boundary, which ends up not being much of a problem as the gas density at the top of the domain is low anyway and the SL does not expand as its temperature is kept constant. In addition, results of our isothermal runs are still well reproduced by simulations with very short cooling time that do possess an atmosphere on top of a denser layer of freshly accreted fluid (\S \ref{sect:with_cooling}).


\section{Results: accretion onto white dwarfs}
\label{sect:WDaccretion}


When presenting results of our simulations we mainly focus on the runs with the WD BCs, i.e. assuming deep ocean of previously accreted matter which can efficiently absorb waves excited in the shear layer. We start by exploring the linear phase of the sonic instability in \S \ref{sect:linphase}, and then discuss in \S \ref{sect:WDisotherm}-\ref{sect:with_cooling} the long term evolution of the SL. In doing this we pay special attention to issues such as the angular momentum transport, wave dissipation and its back-reaction on the SL properties, vorticity generation, mixing, etc. In discussing our results we will often show the behavior of variables, which have been averaged over the $x$-coordinate, i.e. 
$\langle f\rangle_x\equiv x_b^{-1}\int_{-x_b/2}^{x_b/2}fdx $, where $x_b$ is the $x$-dimension of the simulation box and $f$ is the variable in question.

Simulations with the NS BCs are briefly reviewed in \S \ref{eq:NS_BC}.


\subsection{Linear phase of sonic instability}
\label{sect:linphase}


\begin{figure*}
\epsscale{1.}
\plotone{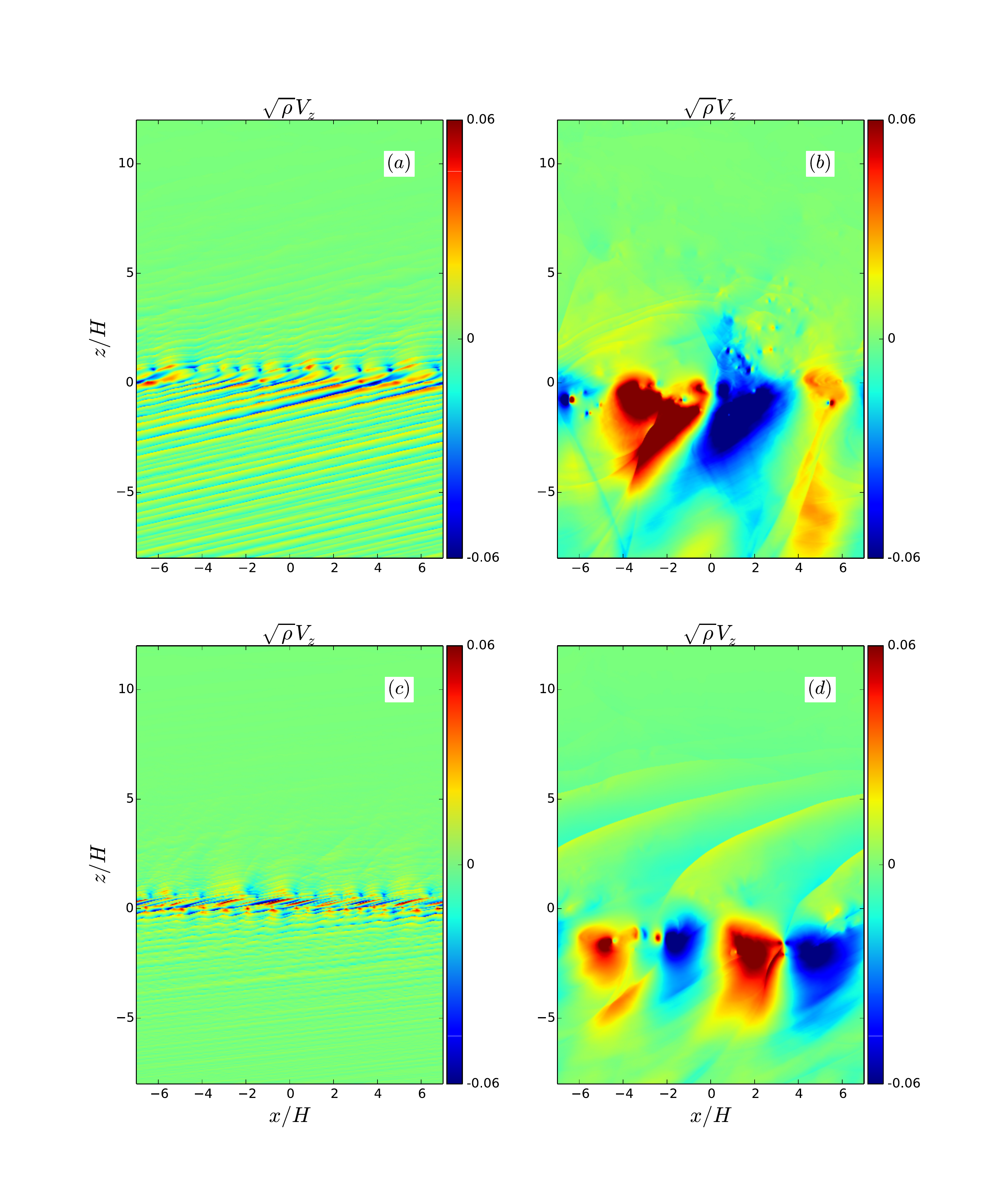}
\caption{Structure of the vertical velocity multiplied by $\sqrt{\rho}$ (to reflect wave action conservation in a vertically stratified medium) in the isothermal simulation during the linear stage of the sonic instability with the WD boundary condition. Simulations with initial Mach number (a) $M=5$, (c) $M=10$ are shown. Supersonic flow in the spreading layer leads to excitation of the non-axisymmetric ($k_x\neq 0$) modes in the shear layer where $\partial v_x/\partial z \neq 0$. Structure of the vertical velocity in the same simulations for (b) $M=5$, (d) $M=10$. during the saturated stage of the sonic instability. \bf{An animation of the $M=5$ flow evolution is available online.}}
\label{fig1}
\end{figure*}

\begin{figure}
\epsscale{1.3}
\plotone{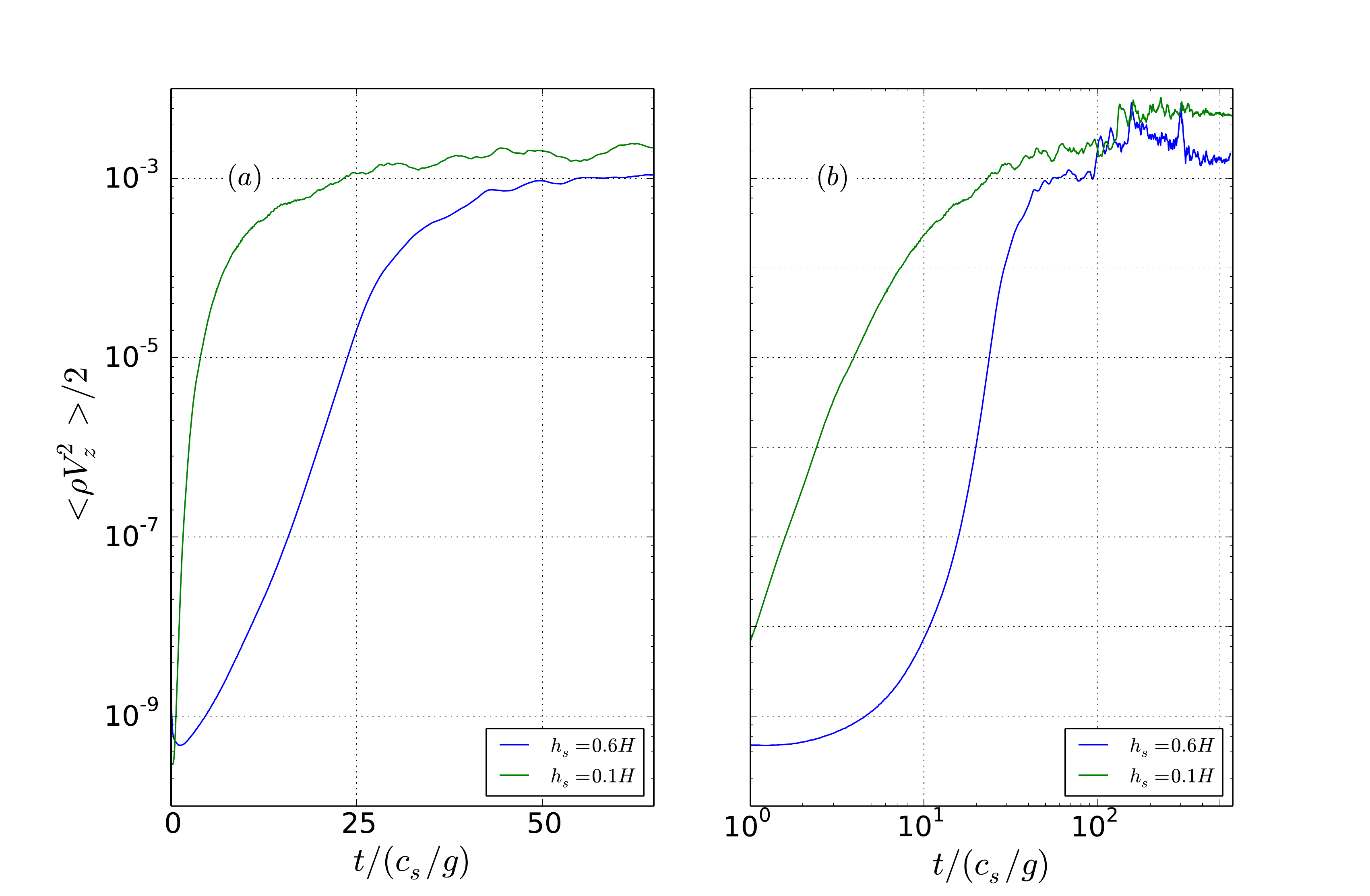}
\caption{The kinetic energy of vertical motions in the isothermal simulation with WD boundary condition for two values of the initial shear layer width: $h_s = 0.6H$ and $h_s = 0.1H$. Panel (a) shows the exponential growth of the instability during the  linear phase in the log-linear scale, and (b) shows the full simulation run in the log-log scale. The saturated values of the vertical kinetic energy are virtually insensitive to the initial shear layer width $h_s$.}
\label{fig19}
\end{figure}

The initial flow has a supersonic drop in the velocity across the shear layer making the region unstable to non-axisymmetric shear instabilities \citep{Glatzel,Belyaev0}. To initialize the instability we put random perturbations to $v_z$ of relative magnitude $10^{-3}$ in the shear layer region. The supersonic shear flow then leads to excitation of the sonic modes, which are non-axisymmetric in global spherical geometry. Their spatial structure in the linear regime in the simulation with isothermal EOS is shown in Figure \ref{fig1}. 

With the free outflow of the excited waves from the shear layer the instability is caused by a radiation mechanism \citep{Belyaev0}. The sonic modes excited in the shear layer propagate both into the star (underlying previously accreted matter) and into the SL with different signs of the wave action. Based on the modal analysis presented in \citet{Belyaev1} we conclude that the most pronounced mode during the linear stage of this simulation is the so-called {\it middle} branch of the acoustic wave. However, the character of the mode can easily change as the simulation evolves in the non-linear regime.

In the absence of dissipation the amplitude of waves propagating into the dense stellar medium decays as $\rho^{-1/2}$ \citep{Belyaev1}, while the z-kinetic energy $\rho v^2_z$ is mostly conserved (see Figure \ref{fig1}). Figure \ref{fig19} shows that, as expected, the kinetic energy of vertical motions (absent in steady state) averaged over the whole box $\langle \rho v^2_z\rangle$ (i.e. over both $x$ and $z$ coordinates) grows exponentially before the sonic instability starts to saturate at $t\approx 30~c_s/g$ (for initial $h_s=0.6H$). This Figure also demonstrates that simulations with smaller $h_s=0.1H$ exhibit larger growth rate of the sonic instability (see also \citealt{Belyaev0}), but converge to almost the same saturated state. This proves that initial shear layer width $h_s$, the value of which may in reality be set by microscopic viscosity, is not important for the saturated state of the sonic instability. We also find the initial behavior to be insensitive to the thermodynamical treatment of the gas. However, the longer-term, non-linear behavior varies for different assumptions about the thermal physics, as we show next.


\section{Isothermal simulations.}
\label{sect:WDisotherm}


To provide direct comparison with the existing results of \citet{Belyaev0} and \citet{Belyaev1} we start by discussing our isothermal runs.

After the sonic instability reaches saturation, properties of the SL evolve on a time scale much longer than the characteristic time $c_s/g$. In this stage sonic waves excited within the shear layer evolve into shocks as they propagate away. Wave dissipation associated with these shocks drives slow rearrangement of the SL properties. 

\begin{figure}

\epsscale{1.45}
\plotone{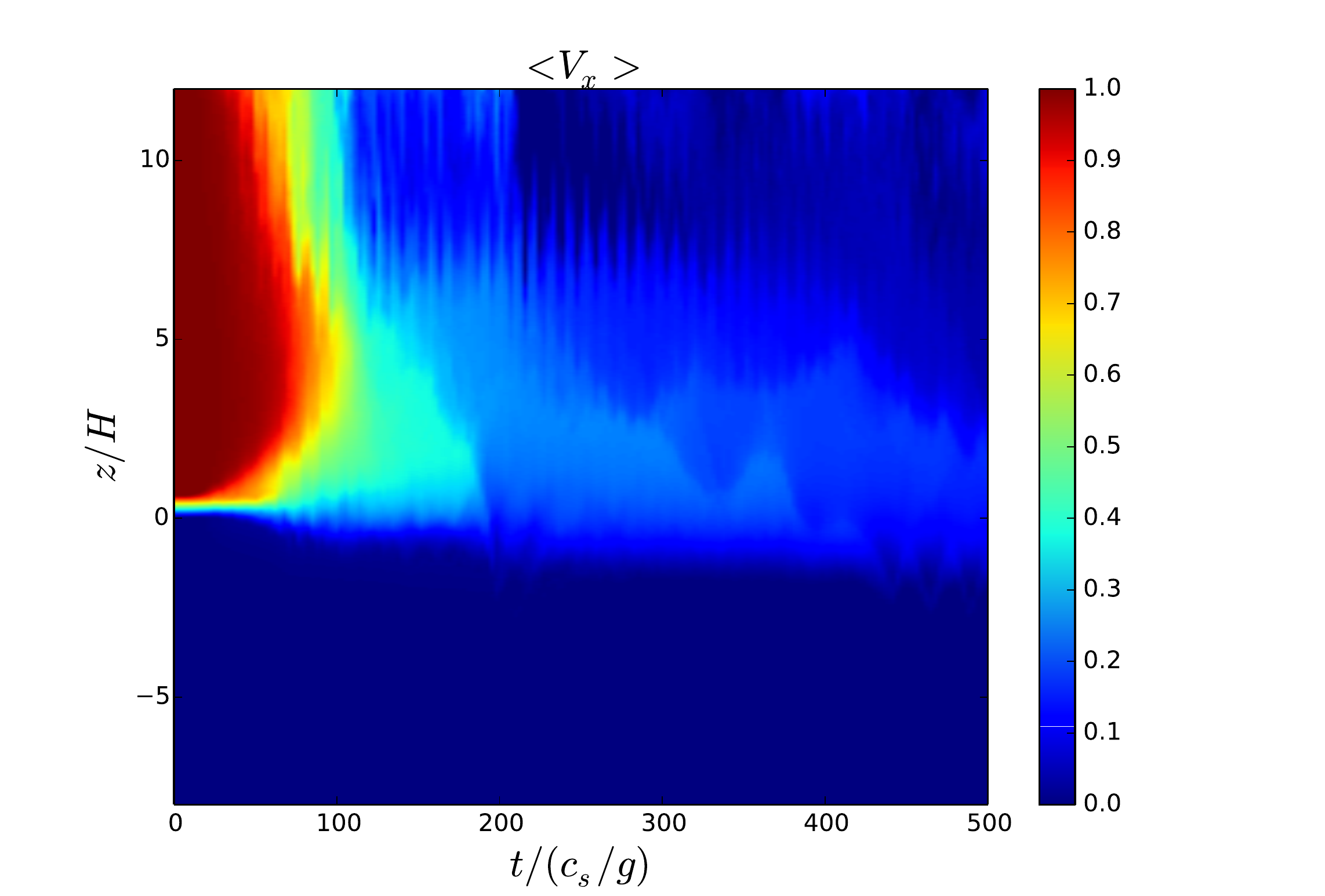}
\caption{Space-time diagram of the x-averaged $\langle v_x\rangle_x$ velocity profile evolution in the isothermal simulation with the WD boundary condtion. A burst of activity at $t \approx 200~c_s/g$ leads to efficient angular momentum transport and rapid velocity drop in the SL. It also results in mixing around the shear layer, see Figures \ref{fig7} and \ref{figs}a. }
\label{fig3}
\end{figure}

Figure \ref{fig3} is a space-time diagram of the $x$-averaged horizontal velocity $\langle v_x\rangle_x$ in $t-z$ coordinates, presenting a detailed view of the horizontal (azimuthal) velocity evolution. In particular, it shows that at $t\approx 60~c_s/g$ vertical oscillations of the shear layer start to develop, which exhibit themselves as vertical striations in this diagram. These oscillations are well described by the dispersion relation for an atmosphere with the isothermal stratification (see \citet{Belyaev2} for description of these modes):
\begin{equation}
\omega^2 = c_s^2 \left(k^2_x + k^2_z + \frac{1}{4H^2}\right) ,
\end{equation}
where $\omega$ is the wave frequency, and $k_x$ and $k_z\approx 0$ are the components of the wave vector. The scale height $H$ stays the same during the instability development in the isothermal case, as the gas temperature does not change. 

\begin{figure}
\epsscale{1.2}
\plotone{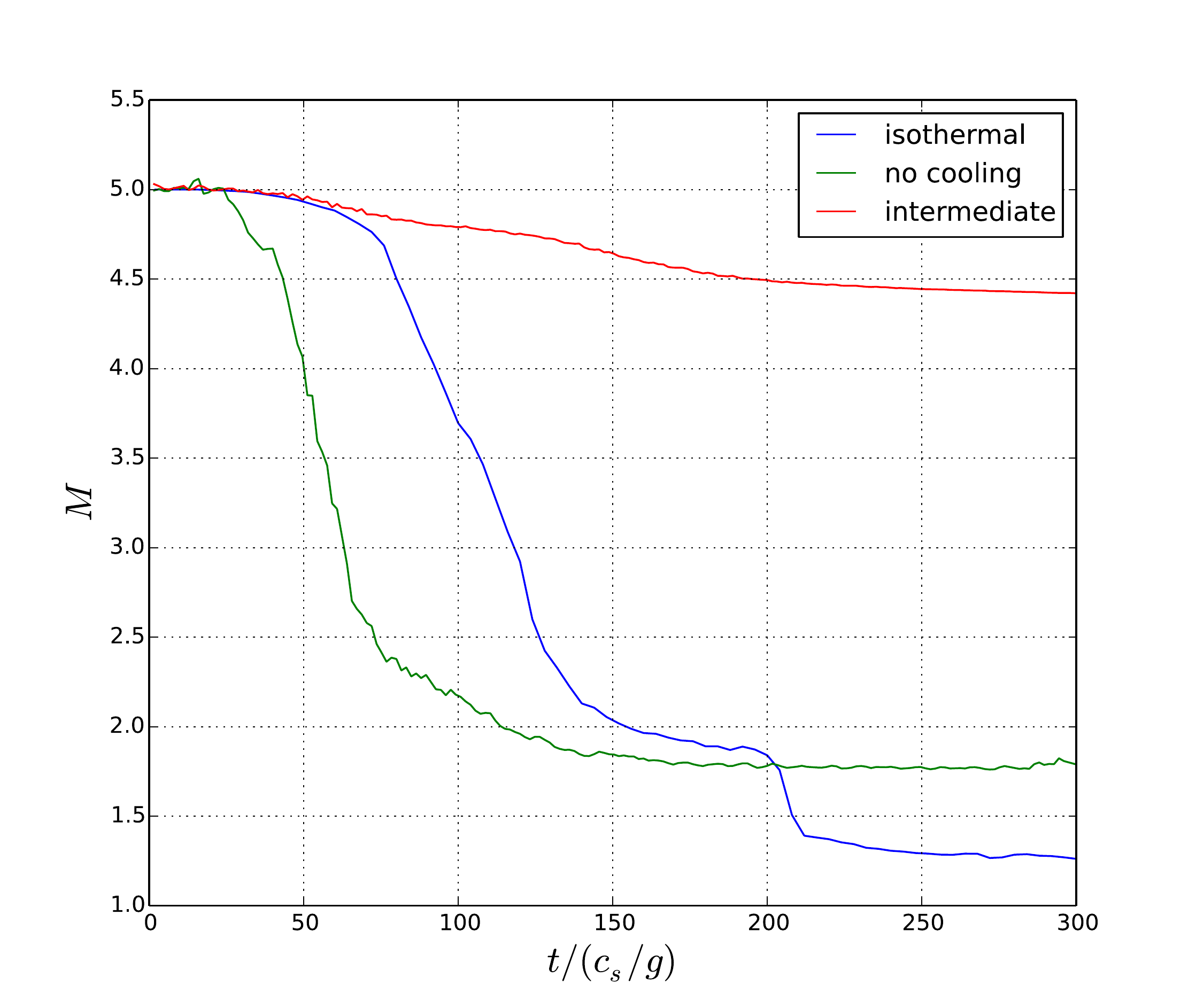}
\caption{Comparison of the evolution of the peak Mach number of the SL (computed with respect to the current, in general evolving, sound speed) in three thermodynamical regimes with WD boundary condition: isothermal, intermediate and no cooling. In the no cooling simulation the Mach number decreases faster, than in the isothermal simulation, because of the temperature increase in the SL region, so that the flow becomes KH unstable at $t\approx 120~c_s/g$. The isothermal simulation becomes KH unstable at $t\approx 200~c_s/g$, after which burst of activity slows down the flow. In the intermediate case, which corresponds to the simulation with cooling time $\tau=0.5 H/c_s$ and is described in Section \ref{sect:interm}, the shocks are weaker that in the isothermal case, and the inward angular momentum transport is less efficient, so the flow does not decelerate enough to become KH unstable.}
\label{fig18}
\end{figure}

On even longer timescales efficient momentum transport between different parts of the simulation domain leads to substantial deceleration of the flow in the SL. As a result, the Mach number of the SL measured with respect to the star (previously accreted fluid at $z<0$) decreases to $M \lesssim 2$. This can be seen in Figure \ref{fig18} illustrating the evolution of the peak Mach number $\langle M\rangle = \mbox{max}\left(\langle v_x/с_s\rangle_x\right)$ in the SL for different assumptions about the thermodynamics. The rapid drop of $\langle M\rangle$ caused by the operation of the well-developed sonic instability at $t\approx 60-120~c_s/g$ turns into a plateau with $\langle M\rangle\approx 1.8$.

Then, at $t\approx 200~c_s/g$ a burst of activity happens (see Figure \ref{fig3}), which drives additional momentum transport and deceleration of the SL flow into the transonic regime, where $\langle M\rangle \approx 1$, see Figure \ref{fig18}. Because of that the interface between the SL and the previously accreted fluid finally becomes unstable to ordinary KH instability \citep{Miles1958}. After this burst of activity and the development of regular KH instability the $\langle v_x\rangle_x$ profile evolves very weakly, see Figure \ref{fig3}. 

\begin{figure}

\epsscale{1.2}
\plotone{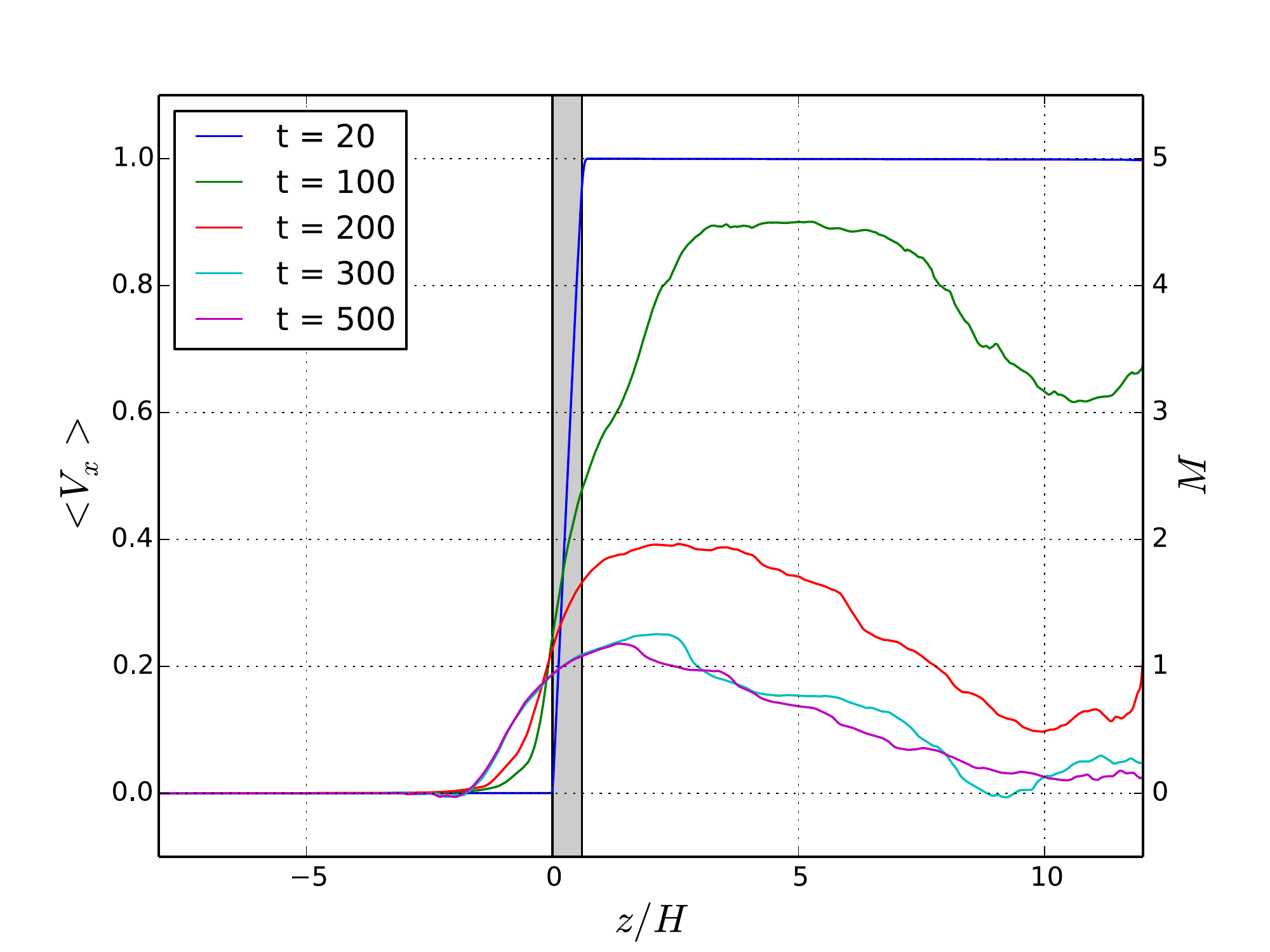}
\caption{Snapshots of the x-averaged transverse velocity $\langle v_x\rangle_x(z)$ profile in the layer at several moments of time in the isothermal simulation with WD boundary condition. Right axis shows normalization with respect to the initial sound speed $M=\langle v_x\rangle_x(z)/c_s$. Grey band shows the initial extent of the shear layer.}
\label{fig4}

\end{figure}

Figure \ref{fig4} displays vertical profiles of  $\langle v_x\rangle_x$ at different times. One can see that momentum in the $x$-direction gets clearly redistributed in the SL --- the uppermost layers of the previously accreted matter (those at $z<0$) get accelerated over time at the expense of the momentum lost by the freshly accreted fluid (at $z>0$). 

The vertical density distribution in isothermal simulations remains effectively unchanged and stays close to the initial exponential profile. This is natural given the lack of back-reaction of wave dissipation on the fluid temperature in the isothermal case, and is different from other thermodynamical regimes that we study later (e.g. \S \ref{sec:adiabatic}). Since the density increases very rapidly with depth, only a small amount of previously accreted gas below $z=0$ is accelerated, as it has enough mass to absorb most of the momentum initially carried by the more rarefied SL above, which has lower inertia.  

\begin{figure}
\epsscale{1.3}
\plotone{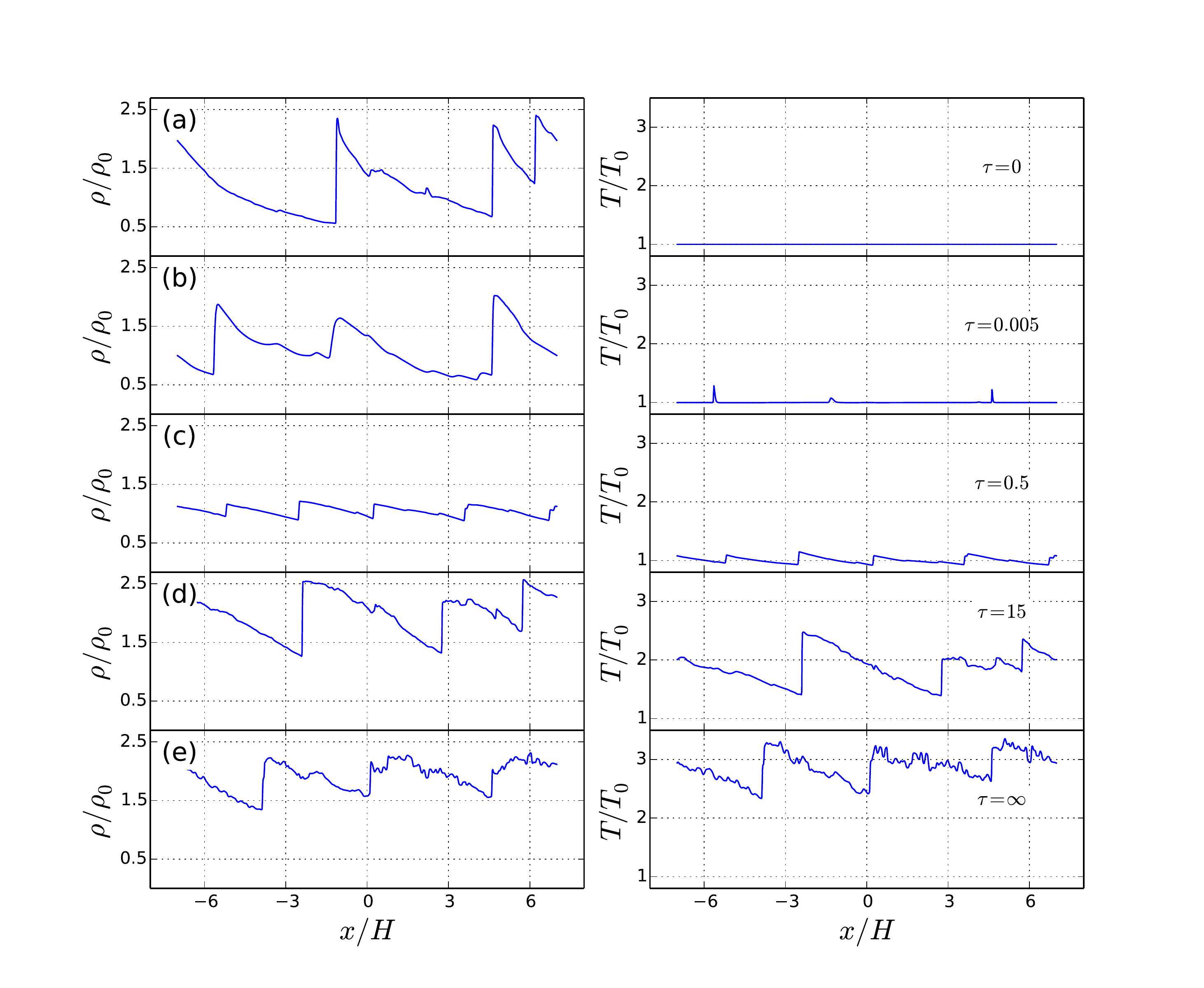}
\caption{Shock structure in our simulations for different cooling times: (a) isothermal, (b) $\tau = 0.005$, (c) 0.5, (d) 15 $H/c_s$ and (e) adiabatic at $t=80~c_s/g, z=0.6~H$. Here $\rho_0$ and $T_0$ are density and temperature values at $z=0.6~H$ and $t=0$. The transition from isothermal strong shocks to intermediate state
  occurs at $\tau \approx 0.1 H/c_s$. This roughly corresponds to the time for the fluid element to pass through two successive shocks. Transition to almost adiabatic behavior happens at $\tau \approx 10 H/c_s$.}
\label{fig16}
\end{figure}

Accretion flows near the BL of a WD are expected to have Mach numbers $M\approx 100$, much higher than $M=5$ that we used in our simulations. In order to explore the role of the flow Mach number, we carried out a simulation with $M=10$. We find that the qualitative behavior of the SL is the same, in both linear (see Figure \ref{fig1}c) and non-linear stages (see Figure \ref{fig1}d) of the sonic instability. In the linear regime, the angle at which the wavefront crosses the x-axis (slope of the "Mach plane") is smaller for larger $M$, $\sin \theta \approx 1/M$ \citep{Belyaev0}. In the non-linear regime we generally observe a wave with a larger horizontal wavenumber (see also \citealt{Belyaev1}). It also takes longer for the instability to saturate for higher $M=10$.

Regarding the late, transonic evolution after the saturation of the sonic instability, it should be kept in mind that this outcome may largely be an artifact of our adopted 2D geometry without explicit forcing of the flow. As mentioned in \S \ref{sect:geometry} our setup does not allow steady state to develop in the system. Nevertheless, the development of sonic modes and their primary role in driving the evolution of the system is still a robust outcome of our simulations.


\subsection{Momentum transport}
\label{sect:mom_transport}


Velocity evolution clearly evident in Figures \ref{fig3} and \ref{fig4} indicates the efficient inward momentum transport in the SL by the acoustic waves. It is driven by the non-linear evolution of the acoustic modes, which ultimately turn into weak shocks and transfer their momentum and energy to the bulk flow.

Figure \ref{fig16} illustrates the structure of shocks into which sonic modes evolve in different runs, including isothermal ones (panel a). It shows the temperature and density slices across the box at $z=0.6~H$, clearly identifying the shock structure. One can see that in the isothermal case shocks are typically rather strong, with density jumps $\Delta \rho/\rho \gtrsim 1$. This results in efficient dissipation and transfer of the momentum and energy carried by waves to the SL fluid. As a result, velocity profile in the SL evolves in time, as  $x$-component of the momentum gets redistributed between the different parts of the simulation box by the wave-driven transport. 

\begin{figure}
\epsscale{1.45}
\plotone{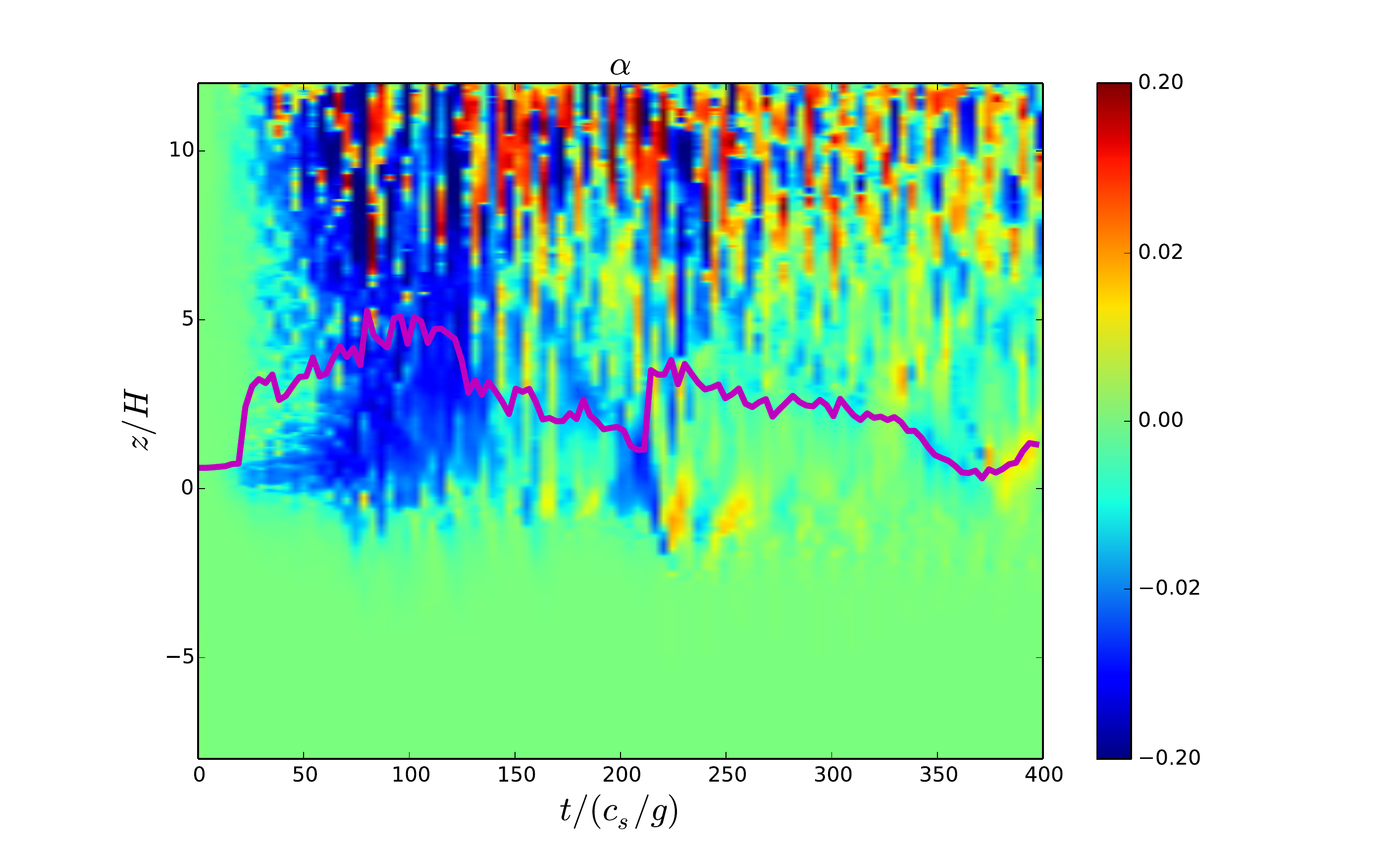}
\caption{Space-time diagram of the x-averaged dimensionless stress $\alpha$ defined by Eq. (\ref{eqalpha}) in the isothermal simulation with the WD boundary condtion. $\alpha$ is negative, thus, the angular momentum is transported inwards, from the SL to the star. The magenta curve shows the height at which $\partial \langle v_x\rangle_x/\partial z = 0$ {changed this} at each moment of time. The fact that stress does not vanish in regions where $\partial \langle v_x\rangle_x/\partial z = 0$ as would be predicted by local models of angular momentum transport, strongly suggests that angular momentum is transported by waves and is fully global. }
\label{fig2}
\end{figure}

Following \citet{Belyaev2} we quantify the wave-driven angular momentum transport via the $\alpha$ parametrization:
\begin{equation}
\alpha = \frac{\langle\rho v^\prime_x v^\prime_z\rangle_x}{\langle p\rangle_x},
\label{eqalpha}
\end{equation}
where $p$ is gas pressure, $v^\prime_i \equiv 
v_i - \langle v_i\rangle_x$ and $\langle \rho v^\prime_x v^\prime_z\rangle_x$ is the momentum flux carried by waves, or Reynolds stress. Figure \ref{fig2} shows a space-time diagram of $\alpha$, which clearly has rather complex structure.

During the linear phase of the sonic instability, for $t\lesssim 40~c_s/g$, momentum transport is very weak. However, after the instability reaches saturation, $\alpha$ takes on large negative value. Because the rotation profile rises at the interface between the star and the SL, the angular momentum is transported {\it inward}, spinning up upper layers of the star (see Fig. \ref{fig4} for the velocity profile). This explains the predominantly negative sign of $\alpha$ until the burst of activity at $t\sim 200~c_s/g$. 

Oscillations of the shear layer lead to alternating direction of the transport transport in the uppermost, rarefied regions of the SL, further complicated by the wave reflection off the upper boundary of the simulation box. However, the transport averaged over time is small. At late stages, when the SL becomes transonic, the value of $\alpha$ decreases and momentum transport slows down.

\begin{figure}

\epsscale{1.5}
\plotone{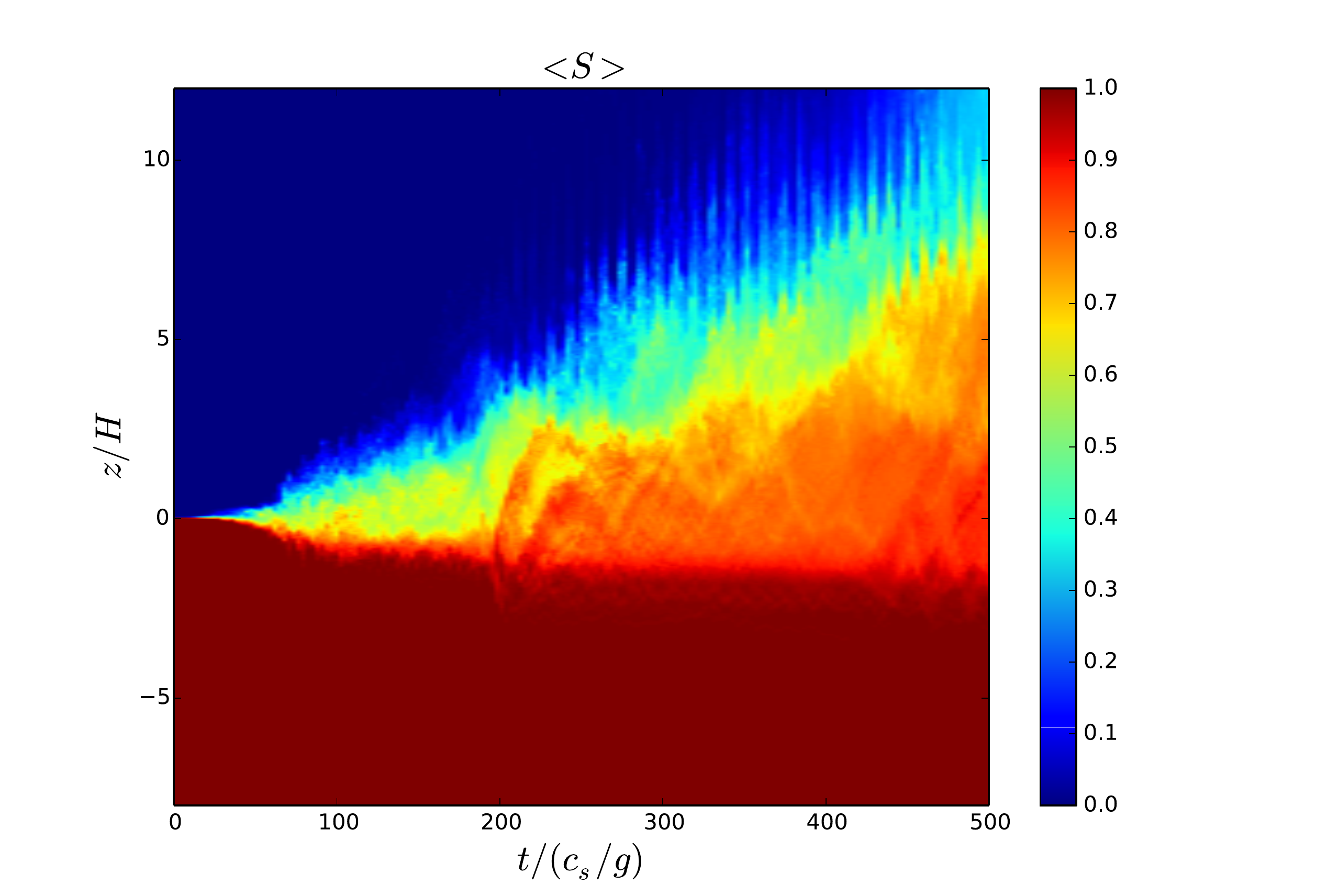}
\caption{Space-time diagram of the x-averaged passive scalar distribution in the isothermal simulation with the WD boundary condition. Mixing accelerates at $t \approx 200~c_s/g$, driven by the burst of activity.}
\label{fig7}

\end{figure}

Wave-mediated transport is intrinsically non-local: the wave has to travel a certain distance, before it shocks and the momentum and energy it carries are dissipated into the mean flow. This non-locality is manifested in several ways. First, looking at Figures \ref{fig3} \& \ref{fig4} we see that velocity in the SL drops fastest at high altitudes, most distant from the shear layer. This is clearly {not a signature of diffusive} evolution that would have been expected from the standard local shear viscosity, and which would result in $\langle v_x\rangle_x$ monotonically increasing with $z$. 

\begin{figure*}
\hspace*{-4cm} 
\epsscale{1.4}
\plotone{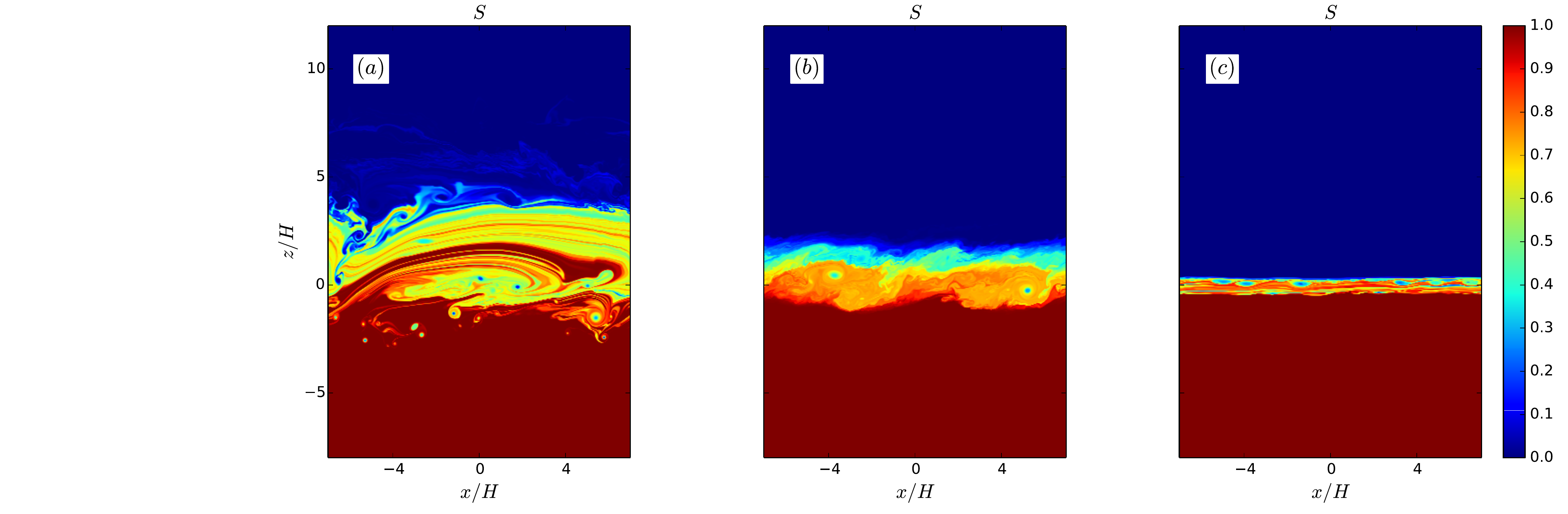}
\caption{Snapshots of the spatial distribution of passive scalar $S$ for different thermodynamical approximations (with the WD boundary condition). (a): At $t \approx 240~c_s/g$ in the isothermal simulation. Mixing is driven by the burst of activity, which strongly decelerates the flow into the transonic regime. The oval structures (cat's eyes) with wispy arms are indicative of the KH instability. (b): At $t \approx 160~c_s/g$ in simulation with no cooling. One can again see the cat's eyes indicative of the KH instability. (c): At $t=160~c_s/g$ in a simulation with the finite cooling time, $\tau_{cool} \approx 0.5 H/c_s$. No mixing happens in this regime as the shocks are quite weak, and no turbulence is generated.}
\label{figs}
\end{figure*}

Instead, the upward-propagating waves launched at the shear layer grow in amplitude because of conservation of wave action and shock readily at high altitudes. Negative angular momentum carried by these waves gets absorbed by the bulk flow and efficiently slows down the material at the top of the SL. This explains the non-monotonic velocity profiles in Figure \ref{fig4}, namely decreasing $\langle v_x\rangle_x$ at high $z$.

Second, Figure \ref{fig2} shows that the dimensionless stress does not in general vanish at the locations where $\partial \langle v_x\rangle_x/\partial z = 0$ as would have been predicted by the local models of angular momentum transport with shear viscosity. The magenta curve in this Figure shows the location at each moment of time of the maximum of $\langle v_x\rangle_x(z)$ determined from Figure \ref{fig3}. It is clear that this curve does not coincide with the locations where $\alpha=0$. These arguments strongly suggest that wave-driven transport dominates in the SL and that it has global nature. 


\subsection{Vertical mixing and vorticity generation }
\label{sect:vert_mixing}


Our calculations also allow us to study vertical mixing between the previously accreted gas (at $z<0$) and freshly accreted material in the rapidly moving SL. If the chemical compositions of the two fluids were different (e.g. as a result of nuclear burning of the previously accreted matter as a result of a nova explosion on a WD or a Type-I X-ray burst on a surface of a NS), then mixing might excavate material with chemical composition different from the accreted gas. This can result in peculiar abundance patterns on the surface of accreting object.

\begin{figure}
\epsscale{1.4}
\plotone{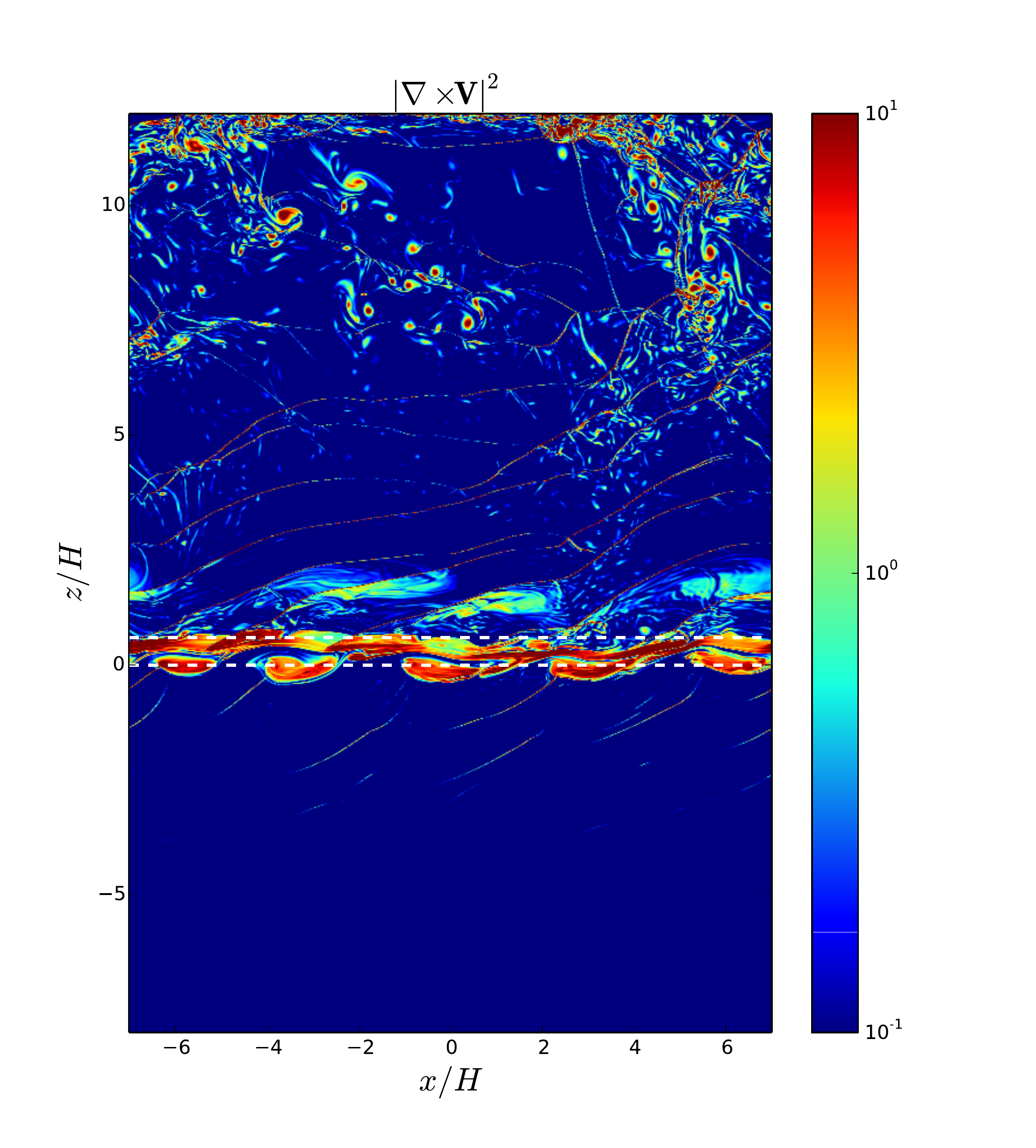}
\caption{Vorticity structure in the isothermal
  simulation with the WD boundary condition at $t \approx 64~c_s/g$. Outside the shear layer (around $z=0$) vorticity is produced only at
the shocks in this simulation. The oscillations of the shear layer
are seen to develop. White dashed lines show the initial extent of the shear layer.}
\label{fig8}
\end{figure}

Figure \ref{fig7} shows the space-time diagram (in $t-z$ coordinates) of the distribution of passive scalar in a typical isothermal simulation. Mixing is essentially absent until sonic instability reaches saturation, but even then the width of the mixed layer increases relatively slowly. The situation changes around $t\sim 200~c_s/g$ when the burst of activity brings the system in the transonic regime, triggering the development of the standard KH instability. The latter drives large-scale vertical motions and turbulence in the box, resulting in efficient vertical mixing. This is illustrated in Figure \ref{figs}a, which shows appreciable propagation of passive scalar into the SL in the post-burst phase, at $t \approx 240~c_s/g$. This map of passive scalar also exhibits oval-shaped vortices. By the end of the simulation shown in Figure \ref{fig7} some previously accreted material has been mixed up to the top of our domain.

Given that mixing is closely related to the generation of vorticity $\omega \equiv \nabla \times \bf {v}$, we also investigate the behavior of this important variable. In particular, Figure \ref{fig8} shows $\omega$ distribution inside the box at $t \approx 64~c_s/g$, after the sonic instability reached saturation but before the burst of activity slows down SL. 

\begin{figure}

\epsscale{1.4}
\plotone{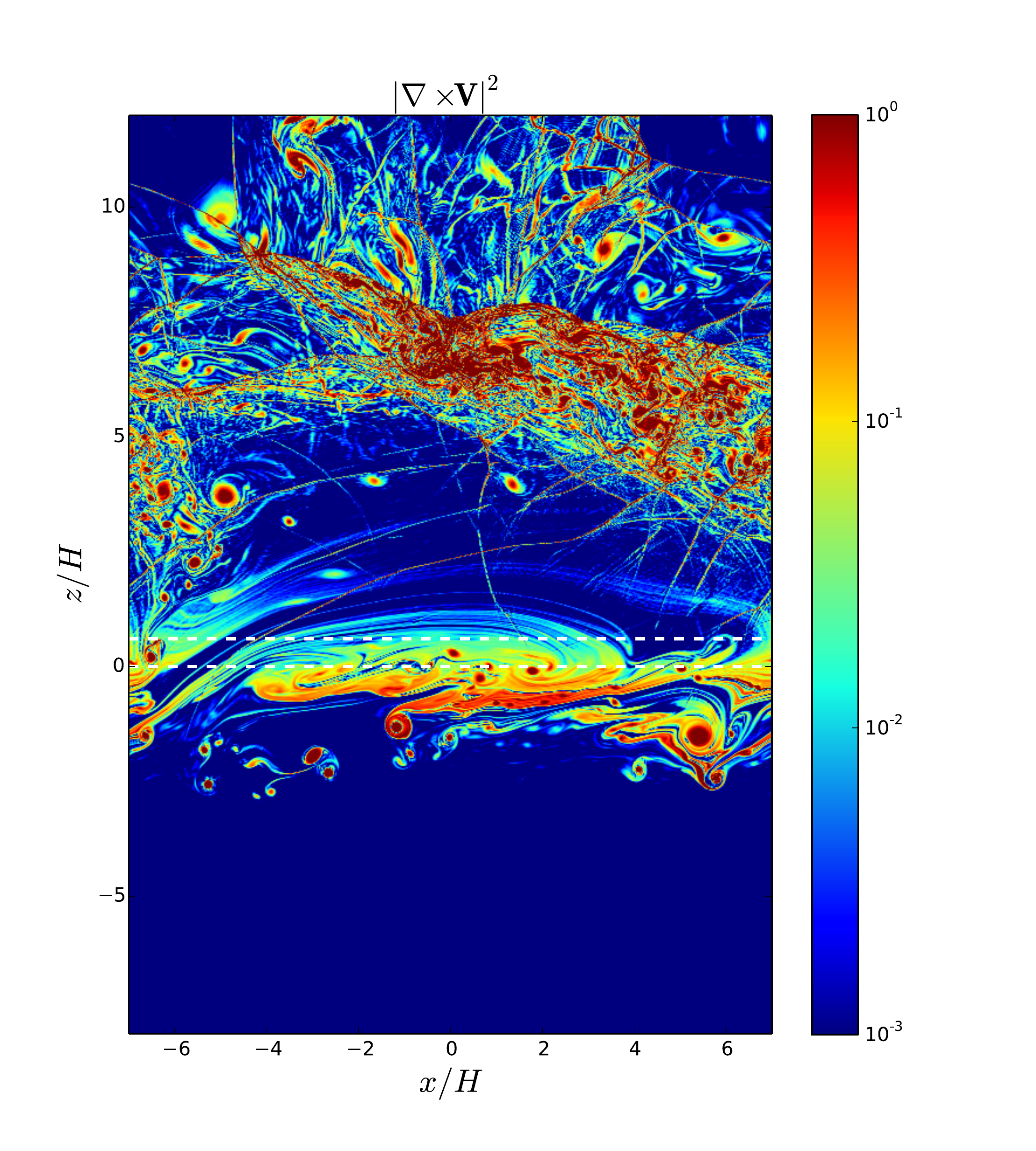}
\caption{Same as Figure \ref{fig8} but at $t \approx 240~c_s/g$. Note the similarity of the vorticity structure at $-5H<z<5H$ with the distribution of passive scalar at the same altitudes in Figure \ref{figs}a. }
\label{fig:vort_iso_late}

\end{figure}

In general, vorticity can be generated both at shocks, and due to nonzero baroclynic $\nabla p \times \nabla \rho$ terms. The latter effect is absent in the isothermal case, and we observe vorticity generation only at shocks (sharp features across the domain), predominantly in the upper part of the box where they are strongest and dissipation is most effective. This map also nicely visualizes  vertical oscillations of the shear layer driven by the sonic instability, which over time result in the diffusion (vertical spreading) of the initially top-hat distribution of $\omega$, see Figure \ref{fig0}e.

Figure \ref{fig:vort_iso_late} illustrates the vorticity distribution at later time $t \approx 240~c_s/g$. One can see that shocks generate significant amount of vorticity in the upper part of the SL, at $z\gtrsim 4~H$ at this time. Also, turbulence caused by the KH instability has completely destroyed the initial narrow band of non-zero $\omega$ near $z=0$ (see Figure \ref{fig0}e). Most of the initial vorticity accumulates in a small number of vortex tubes near the shear layer (saturated round structures). The more diffuse vorticity distribution near the shear layer bears striking resemblance to the distribution of passive scalar at $-2~H<z<4~H$ in Figure \ref{figs}a. This similarity confirms close connection between the mixing and vorticity evolution in the SL.


\section{Simulations without cooling}
\label{sec:adiabatic}


Next we investigate the operation of sonic instability in the regime of no cooling with $\gamma\neq 1$. This is a thermodynamic limit of infinitely long cooling time $t_{cool}=\infty$, opposite to the isothermal case ($t_{cool}=0$) investigated in the previous section. Simulations presented in this section again employ the WD boundary condition, see \S \ref{eq:BC}. It should also be kept in mind that these runs feature a hot atmosphere at the top of the SL.

\begin{figure}
\epsscale{1.4}
\plotone{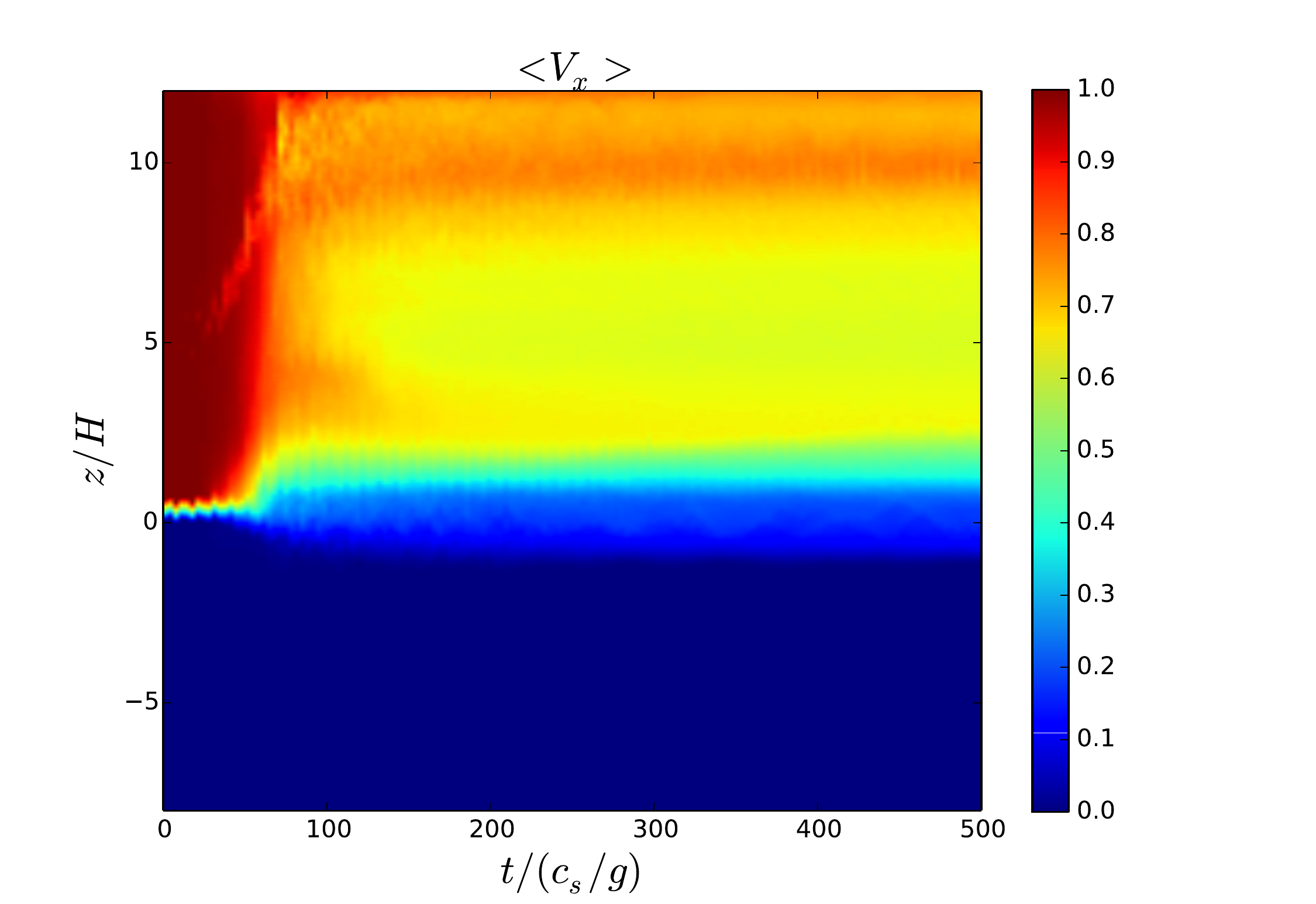}
\caption{Space-time diagram of the x-averaged $\langle v_x\rangle_x$ velocity profile evolution in simulation with no cooling.}
\label{fig111}
\end{figure}

We found that sonic modes are again efficiently excited in the shear layer due to the radiation mechanism, and the linear development of the sonic instability is quite similar to the isothermal case. However, the non-linear evolution with $\Lambda=0$ differs from what is observed in the isothermal case. 

Figure \ref{fig111} shows a space-time diagram of $\langle v_x\rangle_x$ in our simulation with no cooling, and one can easily see the differences with the isothermal case. First, the SL deceleration with time is not as significant as in Figure \ref{fig3}. Second, evolution of $\langle v_x\rangle_x$ profile effectively stalls after $t\approx 80~c_s/g$. Third, the vertical profile of $\langle v_x\rangle_x$ can be roughly described as monotonically rising with $z$, unlike the profiles seen in Figure \ref{fig4}.

\begin{figure}
\epsscale{1.3}
\plotone{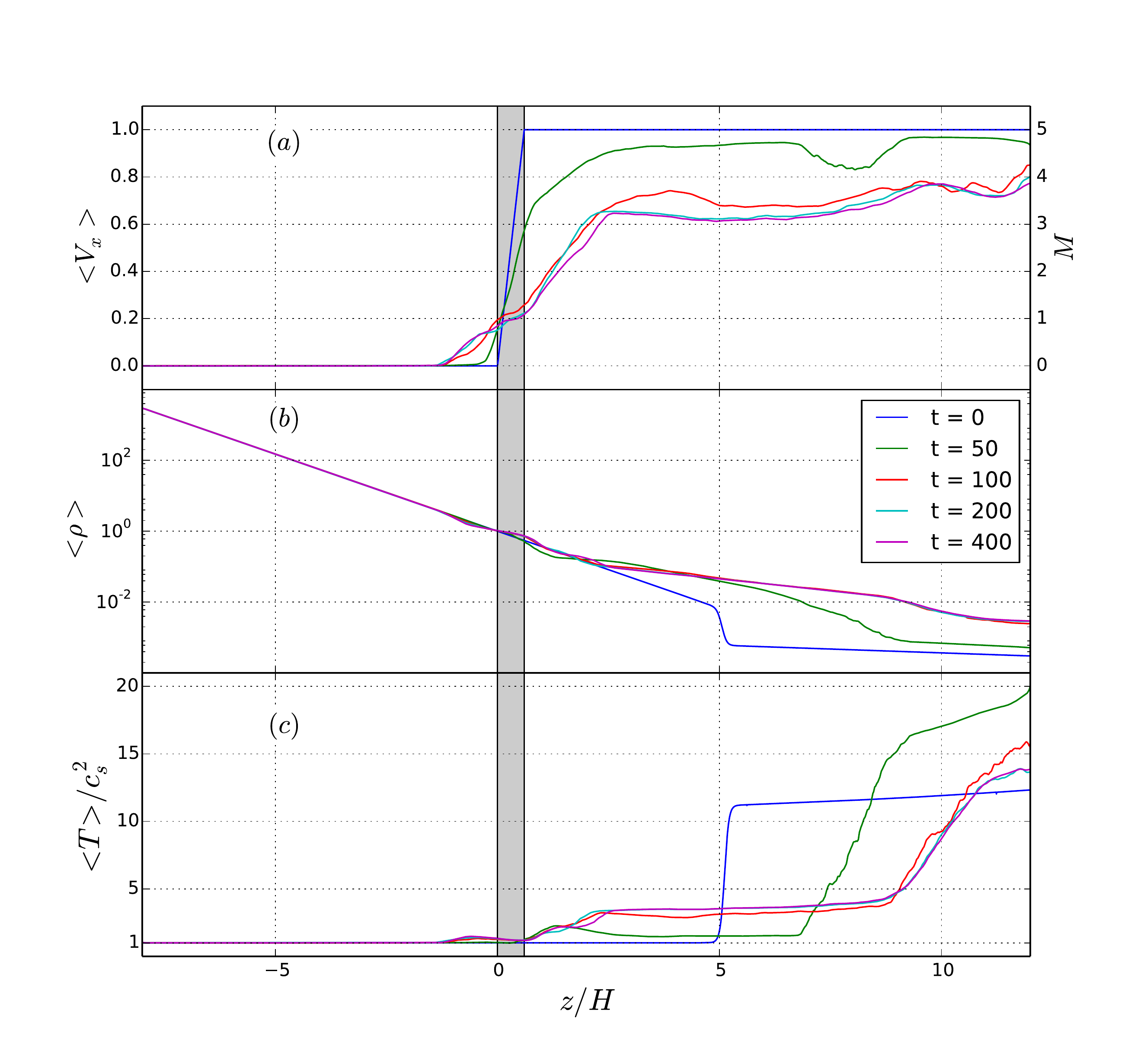}
\caption{Snapshots (times indicated on panel (b)) of different SL characteristics in simulations with no cooling. (a) The x-averaged velocity $\langle v_x\rangle_x(z)$ profile. Right axis shows normalization with respect to the initial sound speed $M=\langle v_x\rangle_x(z)/c_s$  (b) The x-averaged density $\langle\rho\rangle_x(z)$ profile. Because of the heating due to dissipation at the shocks, the SL expands, and the upper atmosphere is pushed to higher altitude, from $z_a\approx 5H$ to $z\approx 9H$. (c): The x-averaged temperature $\langle T\rangle_x(z)$ profile. Temperature evolution results in sound speed increase above the shear layer. The grey band shows the initial extent of the shear layer.}
\label{fig9}
\end{figure}

These points are even better illustrated by Figure \ref{fig9}a, which shows several snapshots of vertical $\langle v_x\rangle_x$ profiles. The extent of stellar spin up, characterized by the extent of the non-zero $\langle v_x\rangle_x$ at $z<0$ is somewhat smaller in this case, simply because the SL does not lose as much momentum as in the isothermal case. 

To better understand the details of this evolution, in Figures \ref{fig9}b and \ref{fig9}c we display snapshots (at the same time) of the vertical profiles of the $x$-averaged gas density $\langle\rho\rangle_x$ and temperature $\langle T\rangle_x$. These Figures clearly show that, unlike the isothermal case, both variables change quite significantly during the evolution. Most of the change happens early on, during the rapid evolution of the SL velocity profile. 

All these changes are ultimately caused by the intense energy dissipation in the SL driven by the damping of the sonic modes, propagating through this region and turning into shocks. Figure \ref{fig16}e  shows that in simulations with $\Lambda=0$ shocks are not as strong as in our isothermal runs (Figure \ref{fig16}a). This is not surprising as the compression ratio for isothermal shocks is proportional to $M^2$, while in the adiabatic case it has a maximum value of 4 for strong shocks. Nevertheless, shocks are still definitely present inside the SL (at $z=0.6~H$) and result in temperature jumps of tens of per cent. These are the same shocks that deliver negative angular momentum flux into the SL, driving evolution of $v_x$. Thus, there is a deep connection between the variation of kinematic ($v_x$) and thermodynamic ($\rho$, $T$) properties of the SL, driven by the wave-induced transport.

Figure \ref{fig9} shows that in the absence of cooling dissipation of acoustic waves rapidly (by $t\approx 100~c_s/g$) increases the temperature of the lower part of the SL (below the hot atmosphere) by a factor of $\approx 3$. This heating of the SL has two important consequences. 

First, the denser part of the SL {\it expands} vertically because of the increased temperature and pressure. Our setup with initial hot atmosphere on top allows the SL to do this, at the expense of shrinking the vertical extent of the atmosphere on top {of it}. As a result, in the end of this run hot atmosphere gets pushed up to $z\approx 9H$, compared to the initial lower boundary of $z_a=5H$. This compression of the atmosphere causes it to heat up beyond the initial $T=10T_0$, as clearly seen in Figure \ref{fig9}c at high $z$. 

Expansion also causes the gas density in the SL to grow, as more mass gets pushed up from the (mildly) heated layers of the previously accreted material at $z\lesssim 0$. Density growth in the hot atmosphere is caused by its compression as we do not allow gas to leave the box through the upper boundary.

Second, temperature growth in the SL dramatically reduces the Mach number of the SL with respect to the star. This can be seen in Figure \ref{fig18}, which shows that Mach number starts to drop precipitously simultaneously with rapid $v_x$, $\rho$, and $T$ evolution, and reaches $\langle M\rangle\approx 1.8$ by $t\approx 160~c_s/g$. Note, that unlike the isothermal case, in which the decrease of $\langle M\rangle$ is caused solely by the decrease of $v_x$, in the case of no cooling $\langle M\rangle$ drops also because of the substantial increase of the sound speed driven by the SL heating.

\begin{figure}
\epsscale{1.4}
\plotone{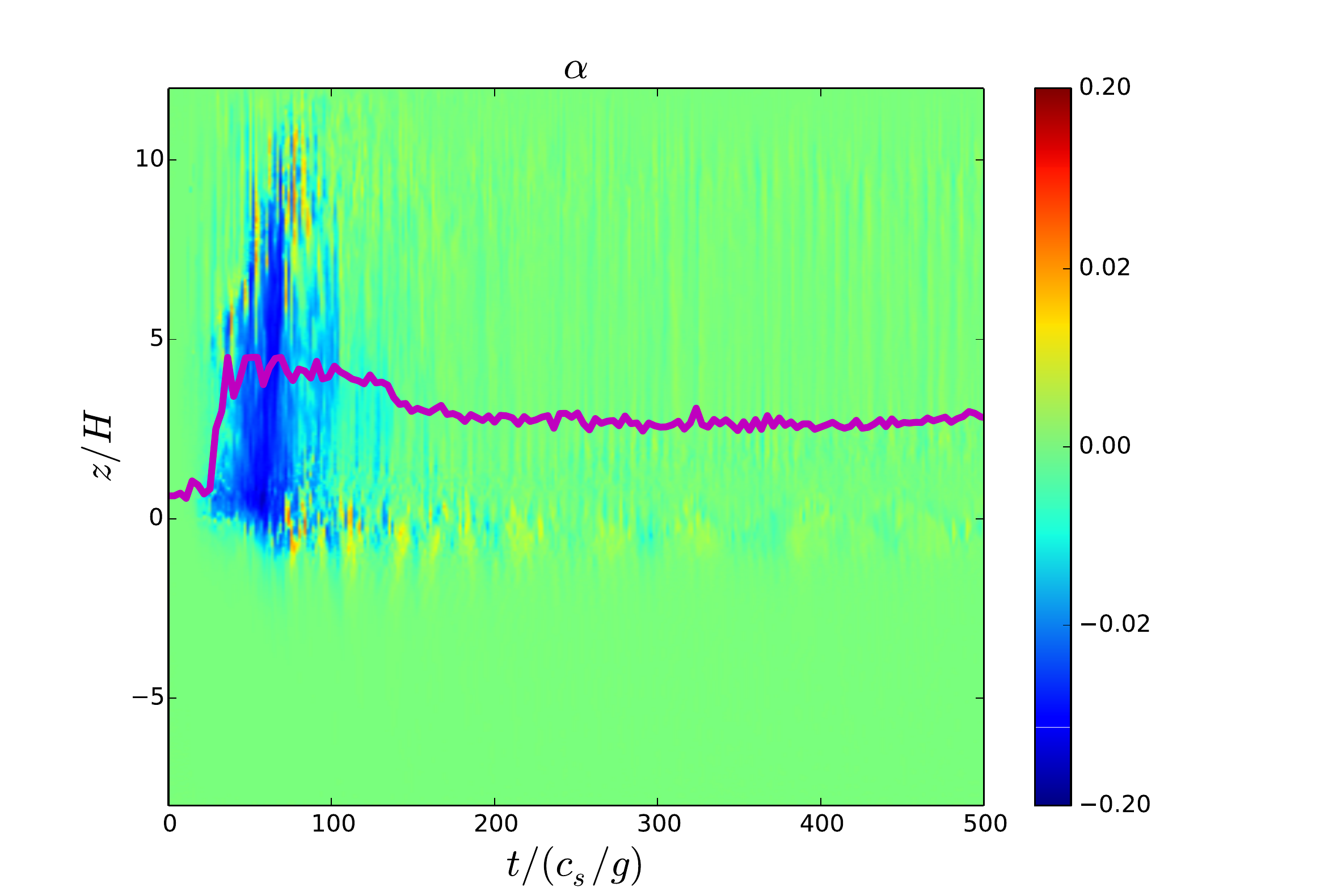}
\caption{Same as Figure \ref{fig2} but for simulation with no cooling. Once again, the stress does not vanish in regions where $\partial \langle v_x\rangle_x/\partial z = 0$, suggesting that angular momentum is transported by waves.}
\label{fig110}
\end{figure}

The reduction of the SL Mach number visible in Figure \ref{fig18} brings it into the transonic regime. As a result, the sonic instability gets quenched around $t\approx 80~c_s/g$, and wave driven transport becomes inefficient beyond this point. This is the reason for the freeze out of the evolution of the SL properties at late times seen in Figure \ref{fig9}. 

Instead, velocity shear still present around $z=0$ drives regular KH instability. At late times we observe KH-like vortices ("cat's eyes") generated in the shear layer, which are clearly visible in the vorticity distributions discussed later in \S \ref{sect:mixing_ad} (see Figure \ref{fig11}). However, the effect of KH instability on the flow is rather local and it does not affect the bulk of the SL (see more in \S \ref{sect:mixing_ad}). 

This general picture is confirmed by the space-time diagram of the dimensionless stress $\alpha$ in Figure \ref{fig110}. It shows that transport is large in amplitude and negative (inward, spinning up the star) only at early times, when the flow in the SL is still supersonic and sonic instability is operating. As soon as the Mach number of the SL drops to about $2$ the transport in the SL gets quenched, halting its evolution. Later KH instability causes only weak transport of alternating sign in the vicinity of the shear layer. 

\begin{figure}
\epsscale{1.4}
\plotone{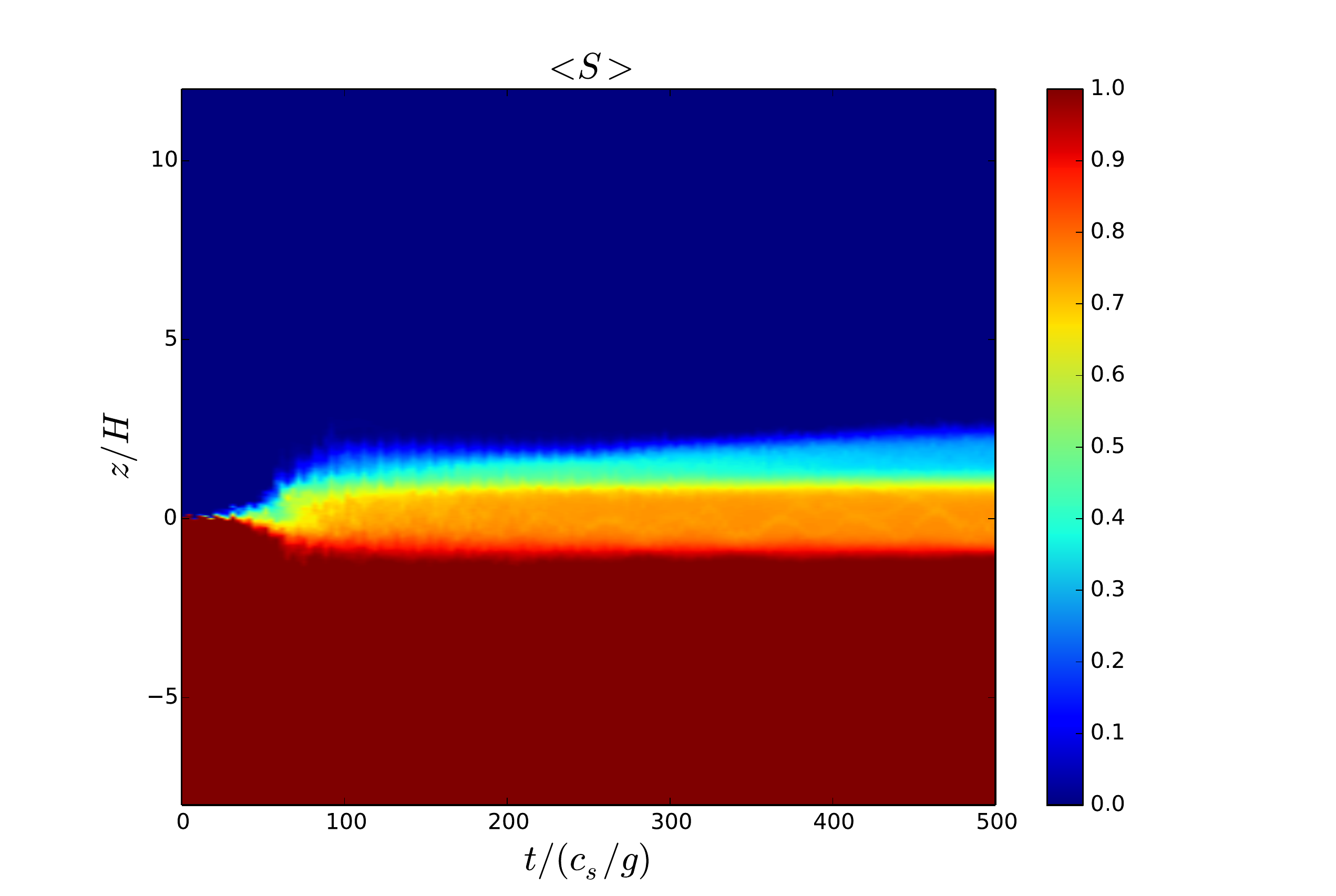}
\caption{Space-time diagram of the x-averaged passive scalar distribution evolution in simulation with no cooling.}
\label{fig112}
\end{figure}

As in the case of the isothermal simulation, the distribution of effective $\alpha$ (see Figure \ref{fig110}) shows that at early times angular momentum transport does not vanish at the locations, where the velocity shear is absent (magenta curve). This once again confirms the global nature of the wave-driven transport, which cannot be captured by the local models with shear viscosity. 

Thus, in the absence of cooling evolution of the SL-star system is self-limited and gets quenched by heating of the SL. In this case a quasi-steady state without much momentum exchange and deceleration of the SL can be reached. However, at fixed latitude this outcome could again be the artifact of our adopted 2D setup (see \S \ref{sect:global_outlook}). The determination of the true steady state will require 3D simulations with radiation transport.


\subsection{Mixing and vorticity generation }
\label{sect:mixing_ad}


Some peculiarities of the SL evolution in the absence of cooling outlined above are reflected in the character of vertical mixing and vorticity excitation. Figure \ref{fig112} is a space-time diagram of the vertical distribution of passive scalar, similar to Figure \ref{fig7}. Mixing is rather weak initially, for $t\lesssim 80~c_s/g$. This is because during this phase the sonic instability operates and associated waves (which are weaker than in the isothermal case) are not effective at vertical mixing of the previously accreted gas into the SL. 

\begin{figure}
\epsscale{1.4}
\plotone{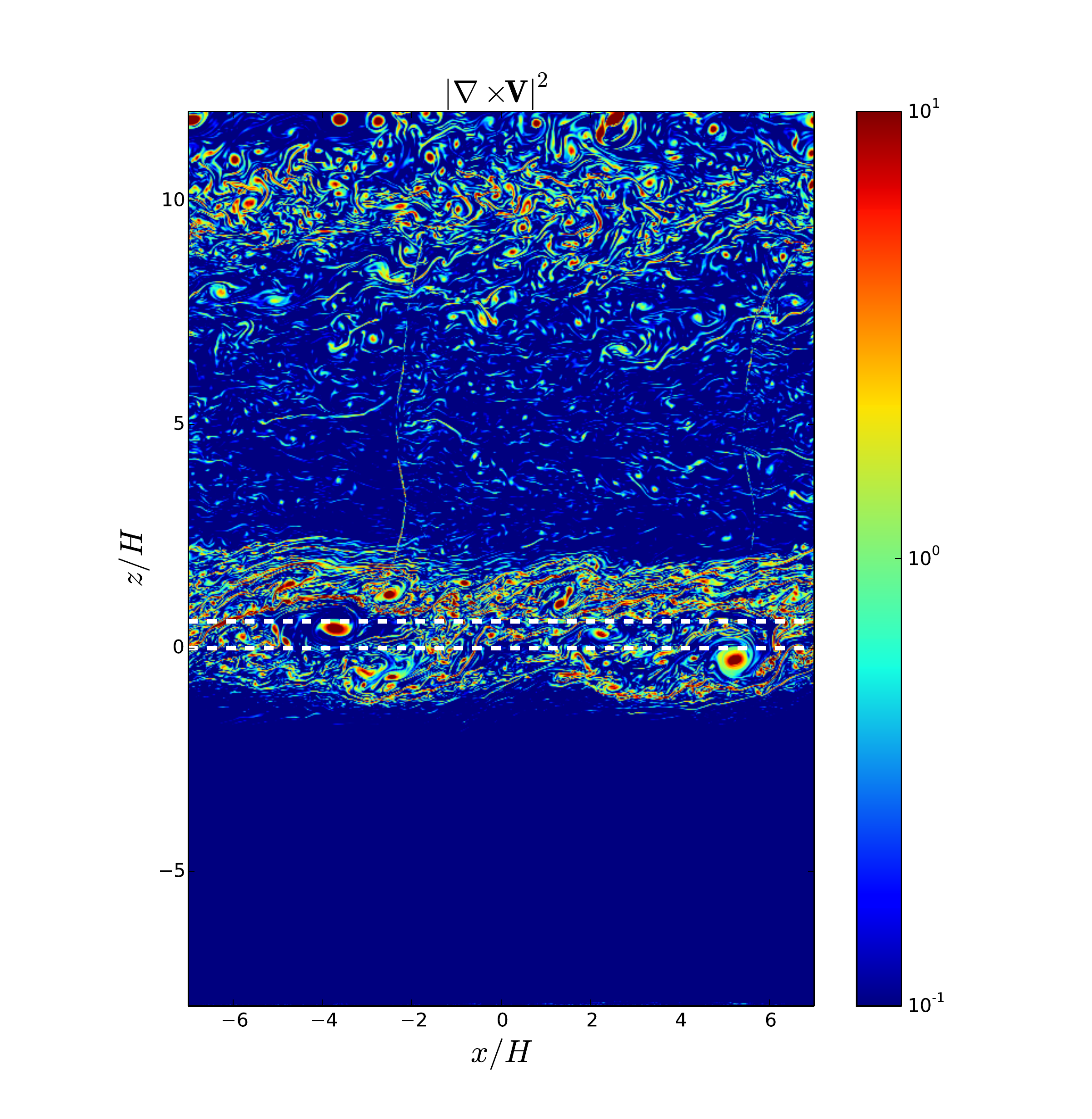}
\caption{Vorticity distribution in simulation with no cooling at $t \approx 160c_s/g$. One can again see the cat's eyes, which are indicative of the KH instability.}
\label{fig11}
\end{figure}

As the Mach number of the SL drops and KH instability gets excited, there is a burst of mixing around $t\approx 80c_s/g$ associated with its development. The spatial distribution of the passive scalar shown in Figure \ref{figs}b has clear signatures of being driven by the KH instability --- "cat's eye" morphology of the vortices driven by the instability. However, this mixing is localized to the immediate vicinity of the shear layer and does not extend into the bulk of the SL. This is in line with the lack of momentum transport in the SL beyond this point in time, which is seen in Figure \ref{fig111}.

Vorticity production in the runs without cooling is somewhat different from that in isothermal runs, see Figure \ref{fig11} for $\omega$ snapshot at $t \approx 160~c_s/g$. 

\begin{figure}
\epsscale{1.2}
\plotone{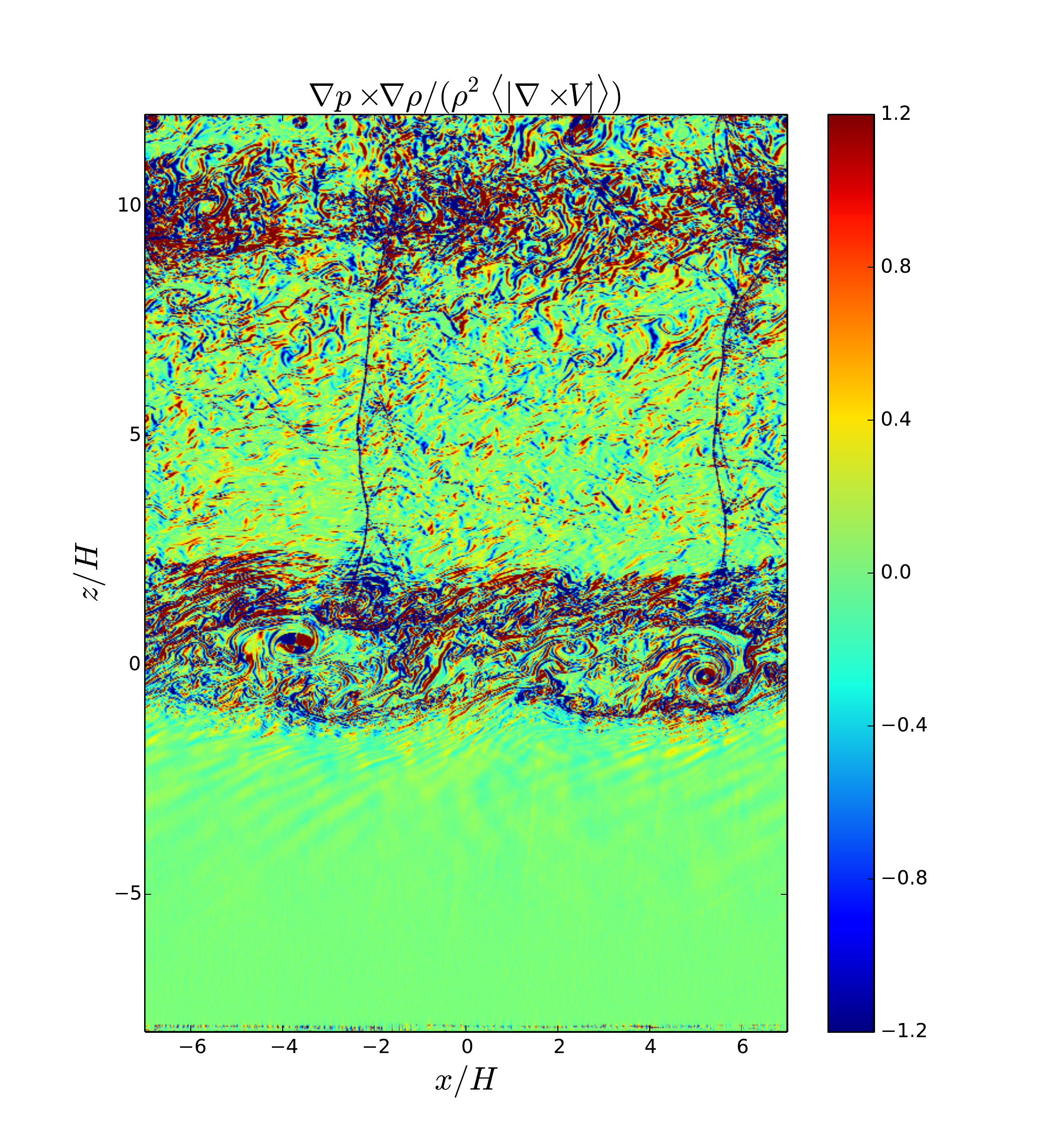}
\caption{Snapshot of the ratio ${\bf {\nabla}} p \times {\bf \nabla}\rho/(\rho^2
  \langle|{\bf \nabla} \times v|\rangle)$ characterizing the baroclynic vorticity production in a simulation with no cooling (and WD boundary condtions).  Here $\langle|{\bf \nabla} \times v|\rangle$ is the space-averaged vorticity in the box. This map shows that baroclynic vorticity production is effective almost everywhere in the SL region since the KH turbulence
  is excited. The discontinuities on this plot are indicators of shocks.
}
\label{fig12}
\end{figure}
First, $\omega$ generation on shocks is not as effective, given their lower amplitude in runs without cooling. Furthermore, rapid quenching of the sonic instability after the SL transitions into the transonic regime makes subsequent $\omega$ production on shocks subdominant. This is reflected in relatively low values of $\omega$ at intermediate altitudes $2~H\lesssim z\lesssim 8~H$ compared to Figure \ref{fig:vort_iso_late}.

Second, in the non-isothermal runs vorticity can also be produced via the baroclynic mechanism, through the non-zero ${\bf {\nabla}} p \times {\bf \nabla}\rho$ term  (which vanishes in isothermal case) in the vorticity evolution equation. This is illustrated in Figure \ref{fig12}, which shows intense baroclynic vorticity production at the top of the box, near the hot atmosphere, and near the shear layer. As a result, these regions feature higher values of $\omega$ in Figure \ref{fig11}. Turbulent mixing of vorticity in the shear layer by the KH instability at late times is very similar to vertical mixing in the same region seen in Figure \ref{figs}b.

\begin{figure}
\epsscale{1.2}
\plotone{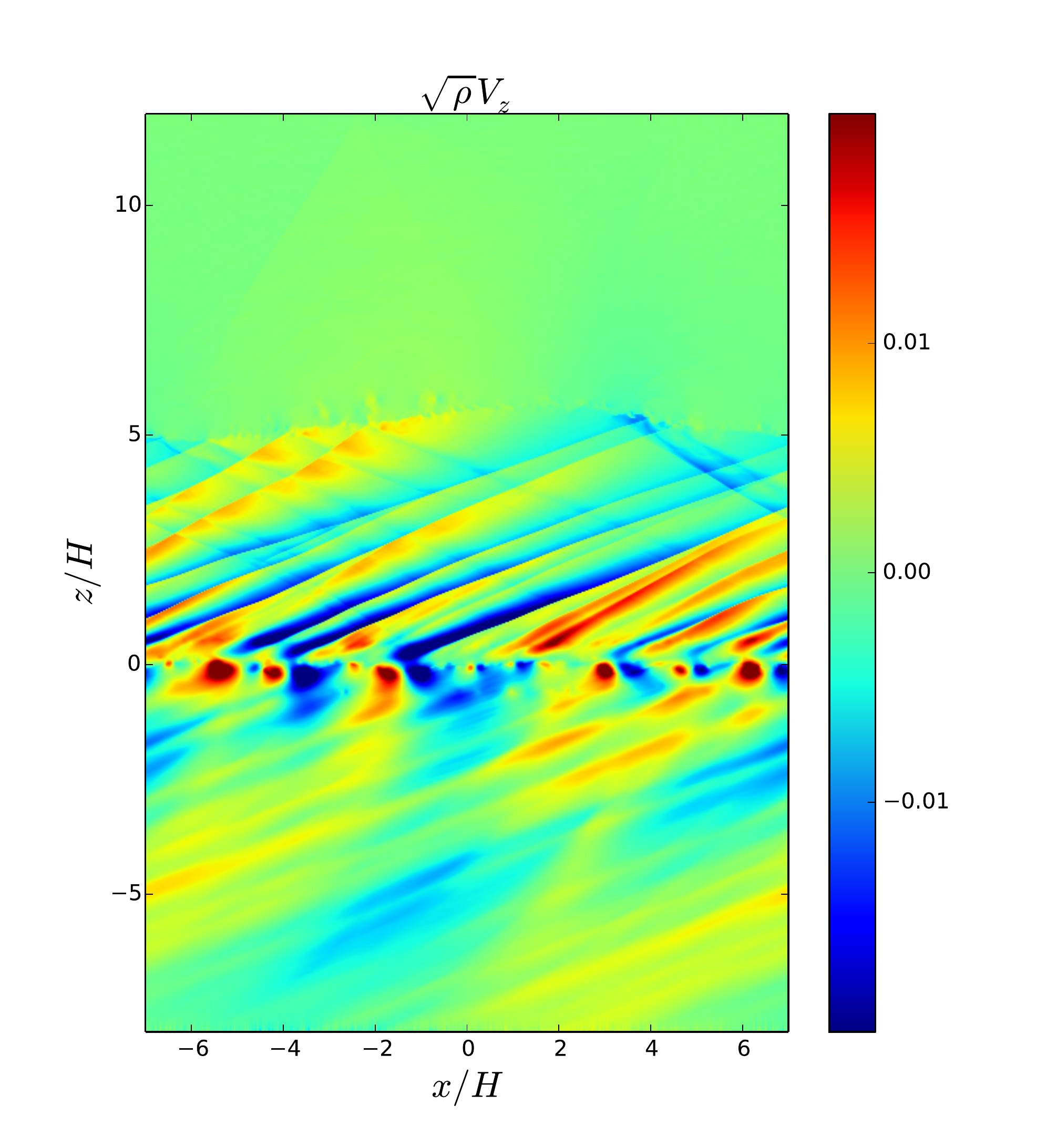}
\caption{Snapshot of $\sqrt{\rho}v_z$ illustrating the structure of acoustic modes in a simulation with cooling time $\tau_{cool} \approx 0.5c_s/g$ at $t\approx 160c_s/g$. At this time the instability have reached saturation and one can clearly see that its morphology corresponds to the middle branch of acoustic modes \citep{Belyaev0}.}
\label{fig:interm_vz}
\end{figure}


\section{Simulations with cooling}
\label{sect:with_cooling}


Having looked at SL in the limiting cooling time regimes of $\tau_{cool}=0$ (in \S \ref{sect:WDisotherm}) and $\tau_{cool}=\infty$ (in \S \ref{sec:adiabatic}), we now present results of simulation with finite cooling time. We run the simulations with a cooling function (\ref{eq:cool}) and $\tau_{cool}$ ranging from $~10^{-3}H/c_s$ up to $50 H/c_s$. This allows us to explore the transition from quasi-isothermal regime at short cooling times, to quasi-adiabatic when $\tau_{cool}$ is long.

The qualitative dependence of the SL evolution upon $\tau_{cool}$ can be understood by looking at the shock structure for different cooling times, displayed in Figure \ref{fig16}. For very short cooling time $~10^{-3}H/c_s$ shocks are almost isothermal with tiny temperature variations and significant density jumps (see panel b). 

\begin{figure}
\epsscale{1.2}
\plotone{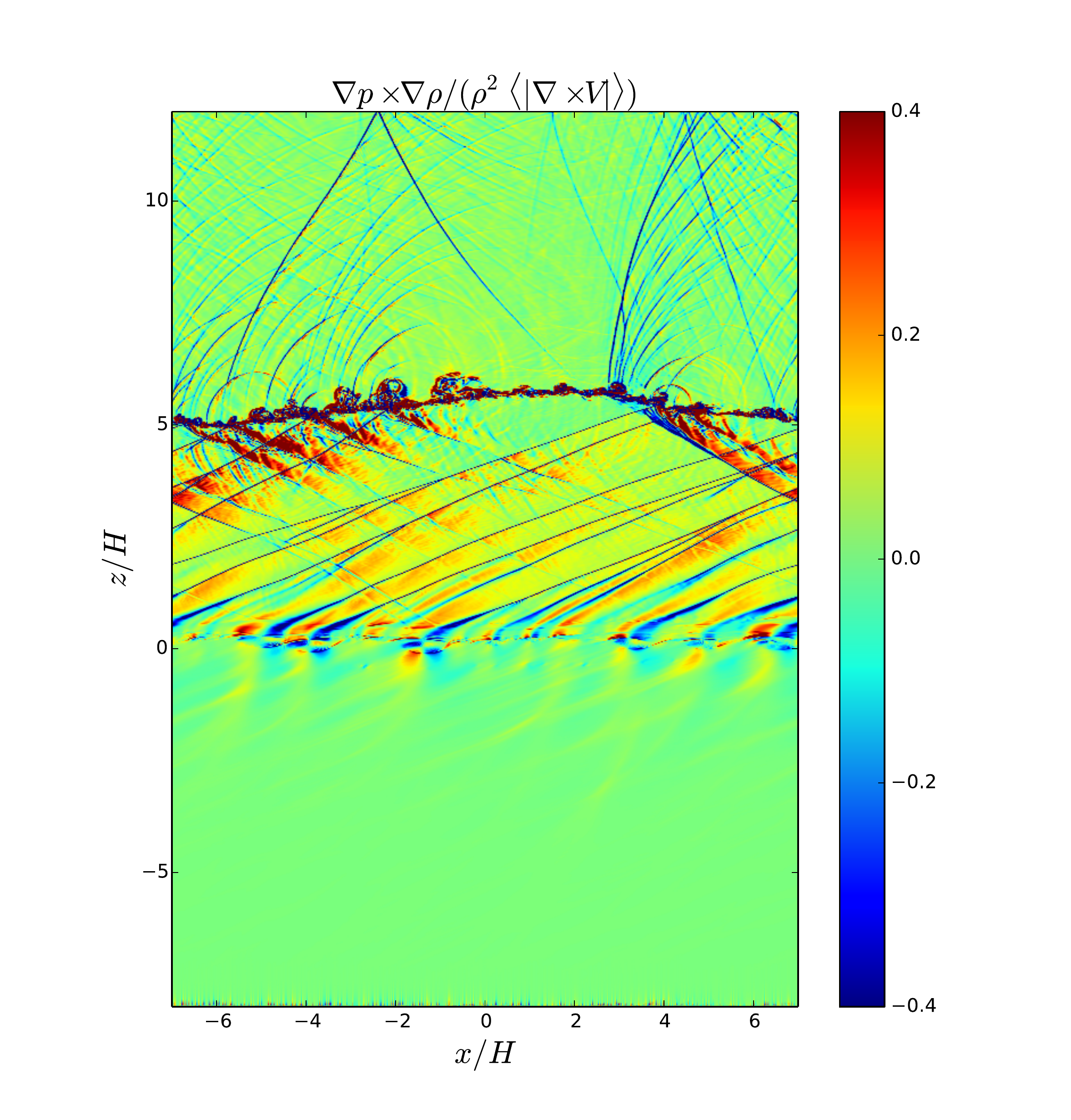}
\caption{Same as Figure \ref{fig12} but for a simulation with cooling time $\tau_{cool} \approx 0.5c_s/g$ at $t=400c_s/g$. Note very well pronounced shock structure in the SL, shock penetration into the upper atmosphere and reflection off the interface between the SL and the atmosphere.}
\label{fig14}
\end{figure}

At longer  $\tau_{cool}\sim H/c_s$ the time for the fluid element to pass through two successive shocks becomes comparable to $\tau_{cool}$, which leads to qualitative changes in the SL behavior. In this regime the fluid element heats up at the shock and cools appreciably before passing through the next shock. This makes the flow behaviour very regular, which is reflected in small shock amplitude in Figure \ref{fig16}c. We call this regime  {\it intermediate} and discuss it in more detail in \S \ref{sect:interm}.

Finally, for $\tau_{cool}$ longer than several scale height crossing times $H/c_s$ the system evolution is similar to the case with $\tau_{cool}=\infty$. As described in \ref{sec:adiabatic}, in this regime the SL gets heated by dissipation of acoustic modes in shocks and puffs up. The flow becomes subsonic and rather turbulent due to the excitation of the KH instability. This results in small-scale variations of temperature and density (see Figure \ref{fig16}, panels d and e for the shock structure).


\subsection{Intermediate cooling regime}
\label{sect:interm}


Figure \ref{fig:interm_vz} shows a snapshot of $\sqrt{\rho}v_z$ in the intermediate cooling regime for $\tau_{cool}\approx 0.5H/c_s$ at time $t=160~c_s/g$. One can clearly see acoustic modes propagating both in the SL (predominantly below the atmosphere) and in the layer of previously accreted matter, at $z<0$. The structure of the velocity perturbation is easily identified as corresponding to the middle branch of the dispersion relation of the acoustic mode \citep{Belyaev0}, with $k_z$ being the same both below and above the shear layer. 

In the intermediate regime the shocks are rather weak (weaker than in the isothermal case) but clearly visible, see Figure \ref{fig16}c, and the shock pattern is stable. Dissipation at shocks is easily balanced by cooling resulting in only a weak change of the density distribution in the simulation box. 

Figure \ref{fig14} shows the snapshot of the baroclynic vorticity production taken at the same time as the $\sqrt{\rho}v_z$ snapshot in Figure \ref{fig:interm_vz}. One can see that in the intermediate cooling regime vorticity is generated only at the shocks and at the boundary between the hot atmosphere and the SL. Vorticity production is less efficient in the bulk compared to the adiabatic simulation.

There is essentially no mixing in the intermediate cooling regime, see Figure \ref{figs}c. This is because the shocks into which acoustic modes evolve are relatively weak, and momentum transport is carried mainly via sonic modes, which are not good at mixing. Also, no bursts of activity are seen during the simulation. This is related to the fact that the temperature in the layer does not increase sufficiently because of efficient cooling. As a result, the Mach number of the flow does not go down substantially (see Figure \ref{fig18}), and the secondary KH instability is not excited during our simulation run, unlike the isothermal and adiabatic cases. For this reason no turbulence is generated in the intermediate cooling case and material mixing between the SL and the layer of previously accreted matter is very weak.


\section{NS boundary conditions}
\label{eq:NS_BC}


We also perform simulations with the NS boundary condition mentioned in \S \ref{eq:BC}. In this setup we impose a reflective BC at the lower boundary of the simulation box (at $z=-1~H$), very close to the shear layer, which now extends from $z=0$ to $z=0.6~H$. This setup is supposed to mainly illustrate the effect of a different BC on the operation of the acoustic modes and is not intended to closely emulate the conditions near the condensed surface of the NS (as envisaged in \citet{InogamovSunyaev}) which should be located at much greater depth. With this new BC the simulations are run for the three thermodynamical regimes, i.e. isothermal, intermediate and no cooling, as in the case of the WD BC. 

\begin{figure}
\epsscale{1.35}
\plotone{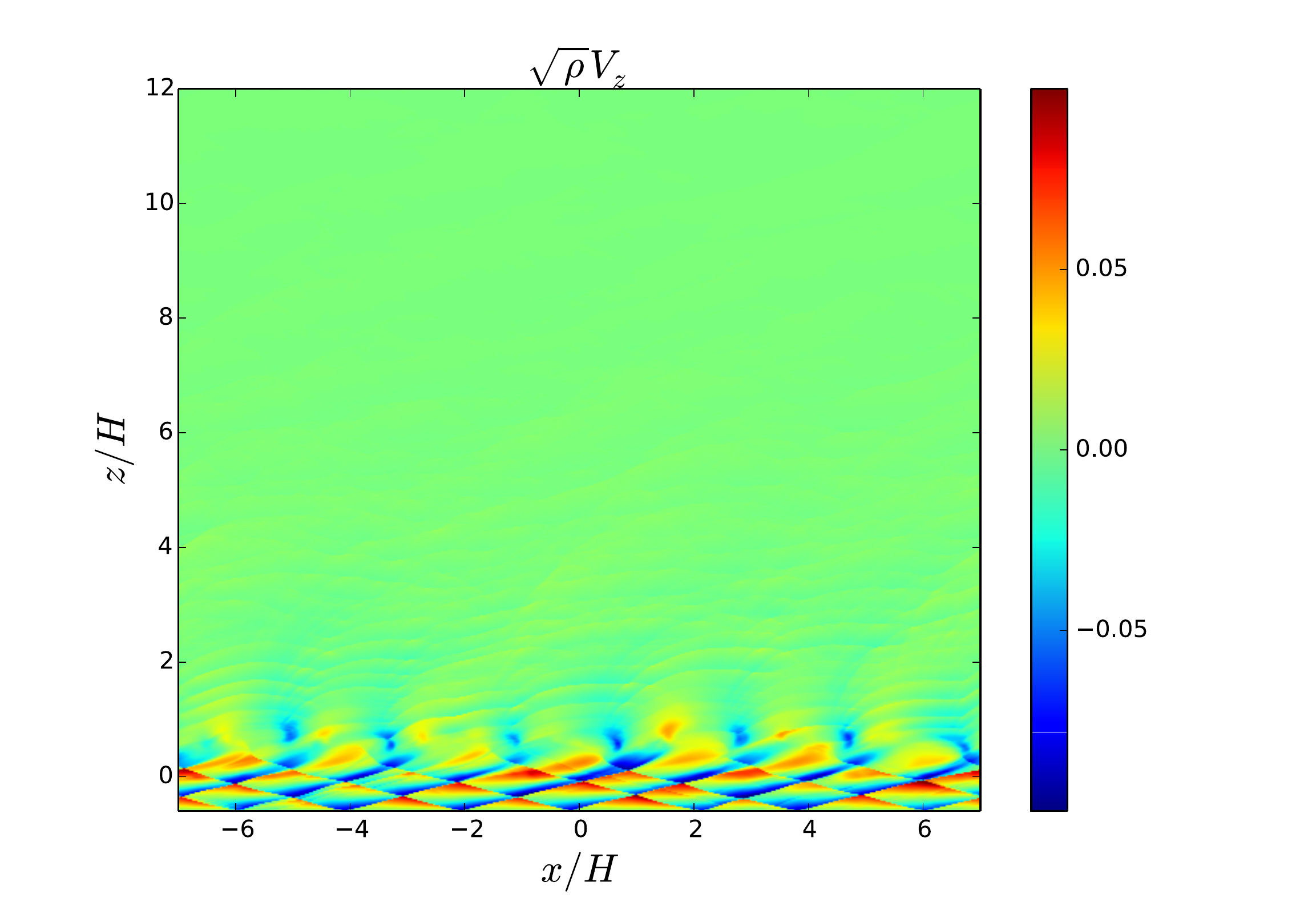}
\caption{Structure of the vertical velocity in the isothermal simulation at $t \approx 20c_s/g$ during the linear stage of the sonic instability with the neutron star boundary condition (NS BC, see text and \S \ref{eq:BC}). Supersonic flow in the SL again leads to excitation of non-axisymmetric modes in the shear layer. The modes become trapped between the lower boundary of the simulation box and the shear layer, and resonate, which accentuates their amplitude.}
\label{fig20}
\end{figure}

We find that there is no qualitative difference with the WD BC, except for the shock structure in the layer of previously accreted material. Figure \ref{fig20} illustrates vertical velocity perturbation at the linear phase of a representative isothermal simulation with the NS BC. Comparing with Figure \ref{fig1} we see the different velocity pattern below the shear layer, with multiple wave reflections off the lower boundary of the box. Similar pattern was observed in \citet{Belyaev0}. The overall mode structure still corresponds to the middle branch of the acoustic instability \citep{Belyaev0,Belyaev2}. However, the instability mechanism is somewhat modified because the downward-propagating modes become trapped between the lower boundary (stellar surface) and the shear layer.

\begin{figure}
\epsscale{1.35}
\plotone{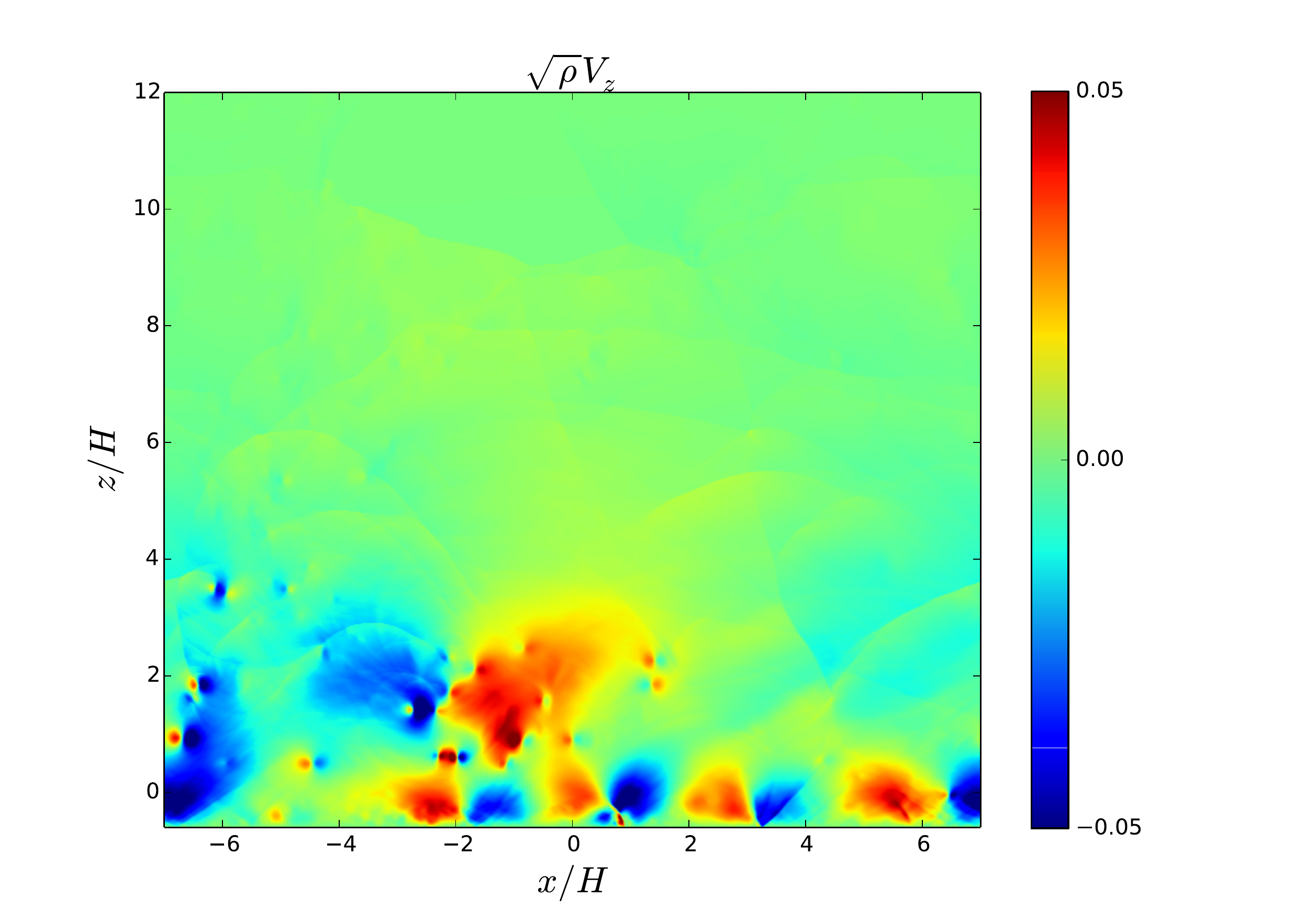}
\caption{Same as Figure \ref{fig20} but at $t \approx 240c_s/g$, illustrating the SL with the NS BC in a saturated state of the sonic instability. At this point the mode pattern is similar to the lower branch of the sonic modes.}
\label{fig:NS_nonlin}
\end{figure}

Figure \ref{fig:NS_nonlin} illustrates the nonlinear phase of the system evolution at $t=240~c_s/g$. One can see that acoustic waves switch to what looks like a pattern characteristic of the lower branch of the mode \citep{Belyaev0,Belyaev2}, with $k_z\approx 0$ in the initially non-rotating part of the simulation domain (at $z<0$). However, the overall behavior of the simulation is still qualitatively similar to the case of the WD BC investigated in \S \ref{sect:WDisotherm}. Because of the efficient angular momentum transport driven by sonic modes and the burst of activity occurring at $t\approx 160c_s/g$ in this simulation, the flow quickly decelerates up to the point when KH instability is excited. This triggers efficient vertical mixing as in the case of the WD BC.

\section{Discussion}
\label{sect:discussion}


Our results clearly demonstrate that supersonic accretion flow over the surface of compact object results in excitation of global acoustic modes. While this has been known in the context of astrophysical boundary layers from the work of \citet{Belyaev1,Belyaev2,Belyaev3}, we have shown this to be true also in the geometry of the spreading layer. We have shown that the sonic modes get excited and efficiently operate with the non-isothermal thermodynamics taken into account and lead to the non-local angular momentum transport. This statement is independent of the boundary conditions used in our simulations, which were designed to replicate either the WD or the NS setup, see \S \ref{eq:BC} \& \ref{eq:NS_BC}.

In our simulations we typically observe excitation of the middle branch of the acoustic mode, which features the same $k_z$ above and below the shear layer, see Figures \ref{fig1} \& \ref{fig20}. Similar behavior was found in \citet{Kley}. The same mode pattern often persists during the saturated phase of the system, after the linear growth has ceased (but see Figure \ref{fig:NS_nonlin} for a counter-example). This can be seen, for example, in Figure \ref{fig:interm_vz}, which shows the system at late times and still exhibits middle branch behavior. This dominance of the middle branch in the advanced stages of the SL evolution is an interesting difference compared to the BL simulations by \citet{Belyaev1,Belyaev2,Belyaev3} and \citet{Kley}, which typically find the lower branch (with vanishing $k_z$ in the non-rotating layer of previously accreted material) to dominate in the saturated phase of the sonic instability. It is not clear if the difference is caused by the local setup of our simulations.

Our analysis provides ample evidence that the angular momentum in the supersonic SL is transported not by the local turbulence, but by acoustic waves excited in the shear layer. As these waves travel away from the shear layer they evolve into shocks due to non-linear effects and dissipate, transferring their momentum and energy to the bulk flow.

Acoustic wave-related momentum transport shows characteristics which are incompatible with the diffusive transport caused by the standard shear viscosity. In particular, we clearly show that stress does not vanish at the locations where $d\langle v_x\rangle_x/dz=0$ (see Figures \ref{fig2} \& \ref{fig110}) as it should in the case of shear viscosity. Instead, there is clear evidence of the truly {\it global momentum transport}, e.g. embodied in the non-monotonic run of $\langle v_x\rangle_x$ as a function of $z$ in Figure \ref{fig4}. For this reason we do not expect models using the standard local $\alpha$-prescription to describe the stress \citep{InogamovSunyaev,Piro} to accurately capture the details of the momentum, mass, and energy transport in astrophysical SLs. Full description of the transport in the SL must explicitly account for its global nature.

We also provide detailed description of the effects of different thermodynamical treatments (isothermal, finite cooling time $\tau_{cool}$, and no cooling) in our runs. The former and the latter extremes (corresponding to $\tau_{cool}=0$ and $\infty$, correspondingly) show rather rapid (within several hundred $c_s/g$) decay of the mean Mach number of the SL, driving its transition into the transonic regime. This ultimately leads to quenching of the sonic instability and excitation of the regular KH instability, that drives local turbulence in the box.

However, isothermal and no-cooling simulations approach the transonic regime via rather different routes. In the former case, efficient momentum transport by the acoustic waves evolving into rather strong shocks results in rapid slowing down of the SL ($\langle v_x\rangle_x$ goes down by a factor of several, see Figure \ref{fig4}), driving its Mach number towards unity. In the latter case, the weaker shocks into which the sonic modes evolve result in modest reduction of the SL velocity (only by tens of $\%$, see Figure \ref{fig9}a). But, unlike the isothermal case, the SL gets substantially heated by shock dissipation, which increases its sound speed and again drives the mean Mach number to relatively small values, see Figure \ref{fig18}.

Cooling with $\tau_{cool}\sim c_s/g=H/c_s$, comparable to the time between the consecutive shock crossings for a fluid element in the SL, results in dramatically different evolution pattern (see \S \ref{sect:interm}): low amplitude acoustic waves dominate transport through the duration of the simulation, heating of the SL and its deceleration are weak, and the Mach number stays in the supersonic regime, which suppresses the excitation of the KH instability.


\subsection{Material mixing in the SL}
\label{sect:mix}


Elemental mixing is an important aspect of the SL physics, especially in the context of accreting WDs. Observations demonstrate that ejecta of Classical Novae are often strongly overabundant (by factors of $\sim 10-100$) in C, O, Ne, etc. compared to Solar values \citep{Gehrz}. As argued by \citet{Truran}, these abundance anomalies cannot be produced as a result of thermonuclear burning during the explosion and should instead be considered as evidence for efficient mixing between the freshly accreted matter and the higher-$Z$ material dredged up from the core of the underlying WD.

Understanding mixing in the SLs of accreting WDs is also crucial for modeling Type Ia supernovae (SN Ia) in the framework of the single degenerate scenario. If the material accreted from normal companion efficiently mixes with the heavier elements dredged up from the CO core of the WD, large amounts of C and O would increase the nuclear burning rates in the freshly accreted layers \citep{Truran82}. This would lead to powerful classical nova explosions, which are believed to eject more material from the WD surface than was accreted prior to that point \citep{Fujimoto,Kahabka}. Thus, in the single-degenerate scenario with efficient mixing the mass of the WD would {\it decrease} as a result of accretion, preventing SN Ia from occurring via this channel\footnote{On the other hand, \citet{PiroWDmix} argued that C/O mixing helps trigger SN Ia via the double detonation channel.}. 

Conversely, it was recently shown by \citet{sn1a} that in the absence of mixing the mass ejected in the thermonuclear runaway is less than the accreted mass, in the wide range of initial masses of WDs and mass accretion rates. This would allow WD to grow in mass until the Chandrasekhar limit is reached, allowing single-degenerate SNe Ia to occur. 

In our simulations we find quite generally that mixing between the SL and the underlying stellar matter is relatively inefficient whenever the transport is dominated by the acoustic modes. Acoustic waves allow efficient exchange of momentum between the different parts of the SL to occur without appreciable exchange of mass and overturning of fluid elements. This tendency is most pronounced in the intermediate cooling case, when the waves are rather weak, see Figure \ref{figs}c. 

However, mixing can become quite intense as the flow decelerates into the transonic regime, leading to the excitation of the KH instability. This instability generates local turbulence, which gives rise to efficient mixing and vorticity production in the vicinity of the shear layer, see e.g. \S \ref{sect:vert_mixing}. Even in this case, however, mixing may not become fully global, see Figure \ref{figs}b.

Because of this difference between the sub- vs. supersonic regimes, mixing of the accreted and stellar material should depend on the thermal physics of the SL, which affects the transition between the two. In isothermal and no cooling cases reduction of the SL Mach number can be quite efficient because of deceleration of the flow and/or SL heating by acoustic wave dissipation, resulting in the early onset of the KH instability and efficient mixing early on. On the other hand, in the intermediate cooling regime (\S \ref{sect:interm}) the wave-driven transport could dominate for very long, suppressing mixing.


\subsection{Energy transport by acoustic modes}
\label{sect:en_transport}


Wave-driven momentum transport inside the SL is necessarily accompanied by the global transport of energy. Previously, \citet{Belyaev1,Belyaev2} have shown that considerable progress in analytical understanding of the wave-mediated transport in the boundary layer setup can be achieved by using linear theory of acoustic mode propagation applicable to weak shocks. Here we attempt to see whether we can reproduce the main characteristics of the energy transport in the SL using this approach. 

\begin{figure*}
\epsscale{1.25}
\plotone{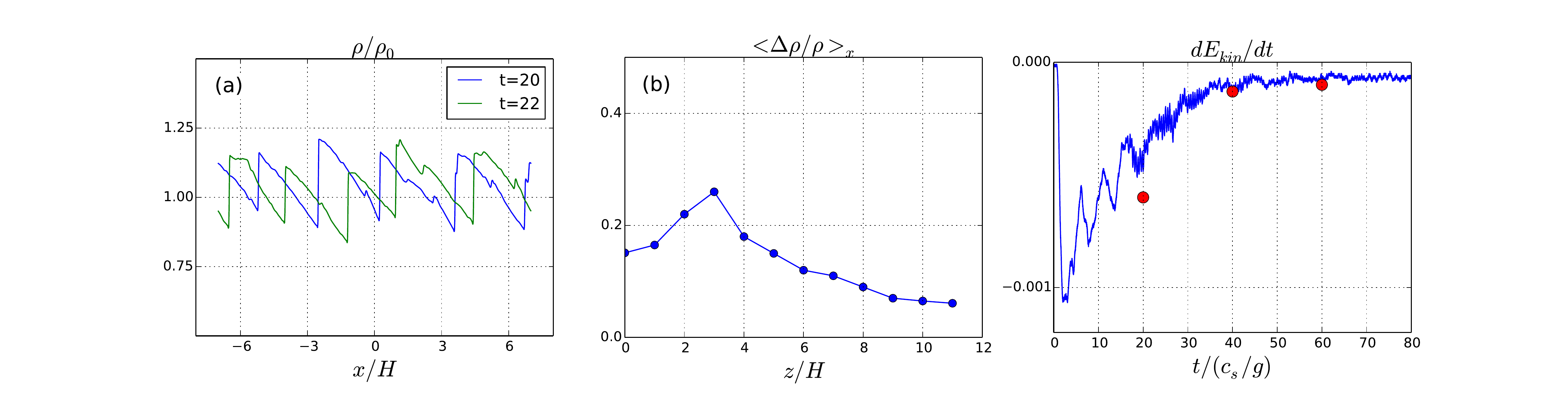}
\caption{Kinetic energy dissipation rate in the intermediate regime with NS BCs: (a) The shock pattern at $t=20~c_s/g$ and $t=22~c_s/g$, from which we infer pattern speed value is $V_p\approx 0.15$. (b) Horizontally averaged shock jump as a function of height within the SL. (c) Kinetic energy dissipation rate as a function of time in simulation (blue curve) and theory (red dots).}
\label{figrate}
\end{figure*}

Since linear theory works best when the shocks are weak and the pattern is stable, we focus on a simulation with intermediate cooling described in \S \ref{sect:interm}, in which the SL state is not perturbed significantly by heating. Using a setup with NS boundary conditions (\S \ref{eq:NS_BC}), we try to compute the rate at which kinetic energy of the flow is irreversibly turned into heat at the multiple shocks seen in Figures \ref{fig:interm_vz} \& \ref{fig14}. For the weak shocks observed in this simulation, the kinetic energy dissipation rate can be found as \citep{Belyaev1}
\begin{equation}
\frac{d E_{\rm kin}}{d t} \approx -\frac{\gamma+1}{6}\int
c_s^2\langle\rho \left(\frac{\Delta \rho}{\rho}\right)^3\rangle N_{s} \left[V_{p} -v_{x}(z)\right] dz,
\label{eqloss}
\end{equation}
to lowest order in $|\Delta \rho /\rho|$. Here $\langle ..\rangle$ implies averaging over individual shocks at a given $z$, $\Delta \rho /\rho$ is the density jump at the shock, $c_s$ is the local sound speed, $N_{s}$ is the number of shock waves in the simulation box, $v_x(z)$ is the actual flow speed and $V_{p}$ is the shock pattern speed. This formula fully accounts for the {\it global nature} of the energy transport by acoustic waves.

We measure $V_p$ by following the propagation of the shock pattern over short interval of time, as demonstrated in Figure \ref{figrate}a; for example, at time $t=20~c_s/g$ we find the pattern speed to be $V_p \approx 0.15$. The shock compression ratio varies relatively weakly through the simulation box as shown in Figure \ref{figrate}b. The vicinity of the shear layer (small values of $z$) contributes most to the integral (\ref{eqloss}) because of weighting by the local density. These measurements allow us to compute theoretical $d E_{\rm kin}/d t$ at a given moment of time, which we compare with the same quantity determined directly from the simulations. 

Results of such comparison are shown in Figure \ref{figrate}c (we compute theoretical $d E_{\rm kin}/d t$ at $t=20,40,60~c_s/g$), demonstrating reasonable agreement between the two approaches. At later times shocks become weaker, and the energy dissipation rate decreases.

The global character of the energy transport in the SL could have important implications for the physics of Type I X-ray bursts in low-mass X-ray binaries (LMXBs) harboring low-magnetic field NSs. Conditions for Type I burst ignition are known to depend on the temperature at the base of the accreted layer \citep{Stro}. This temperature is expected to be high if the deceleration of the SL is due to the some type of shear viscosity, as then the heating is strongest at the base of the layer, where cooling is slowest, resulting in high $T$. It was argued by \citet{IS2} that in this case the conditions for triggering Type I bursts would be difficult to achieve. Instead, contrary to observations, accreted material would always burn stably at the base.  

The situation may easily be different in the supersonic SL once the key role of the wave-driven transport is recognized. Indeed, {\it upward-traveling} waves launched at the base of the SL, in the shear layer, will {\it propagate away} from this layer, carrying to higher altitudes a significant share of the mechanical energy released in deceleration of the flow deep down, just as they do in our simulations. There waves would dissipate there, ultimately releasing their energy as heat, but this would happen {\it closer to the SL photosphere}, at lower depth than that of the shear layer. This sub-surface character of heating could substantially reduce the temperature at the base of accreted layer, possibly providing the conditions necessary for triggering Type I X-ray bursts in LMXBs.   

On the other hand, if {\it downward-propagating} waves carry a significant fraction of the energy and momentum released in the shear layer, their dissipation at some depth within the star may provide a subsurface heating source. This would not only increase the temperature of the object's interior, but also give rise to the heat flux towards the stellar surface, which may have observable consequences. For example, recent observations \citep{MAXI} of the thermal state of the accreting neutron star transient MAXI J0556-332 suggest a large deposition of heat in the shallow outer crust from an unknown source. The additional heat injected is $\approx 4-16$ MeV per accreted nucleon. At the same time, the total energy released in decelerating the (initially almost Keplerian) accretion flow in the SL should provide about $GM_\star m_p/R_\star\sim 100$ MeV per accreted nucleon. Thus, even if only $\sim 10\%$ of this energy gets carried by the sonic waves excited in the SL deep into the outer crust of the NS, where it gets dissipated (via some yet unexplored mechanism) heating the crust, then the global acoustic modes may naturally explain the intense subsurface heating in accreting NSs.  

This discussion shows that the accurate determination of the relative contributions of the upward- and downward-propagating wave energy and momentum fluxes is an important task with a number of observational implications. We leave the detailed exploration of this issue to a future study with more realistic physical inputs.


\subsection{Understanding global SL properties}
\label{sect:global_outlook}


As discussed before (\S \ref{sect:WDisotherm} \& \ref{sec:adiabatic}), our simulations often find rapid evolution of the system, SL deceleration, and departure from the supersonic regime, see Figure \ref{fig18}. This outcome is a direct consequence of our 2D setup without explicit injection of momentum, in which a steady state SL arrangement is not possible. In reality, continuous mass transport in the meridional direction (not captured in our simulations) should drive the system to a quasi-stationary configuration. It is quite plausible that in this configuration the Mach number of the SL {\it at a fixed latitude} would never drop to low values, precluding the regular KH instability from happening and leaving acoustic modes as the only means of the angular momentum transport in the SL.

\begin{figure}
\epsscale{1.15}
\plotone{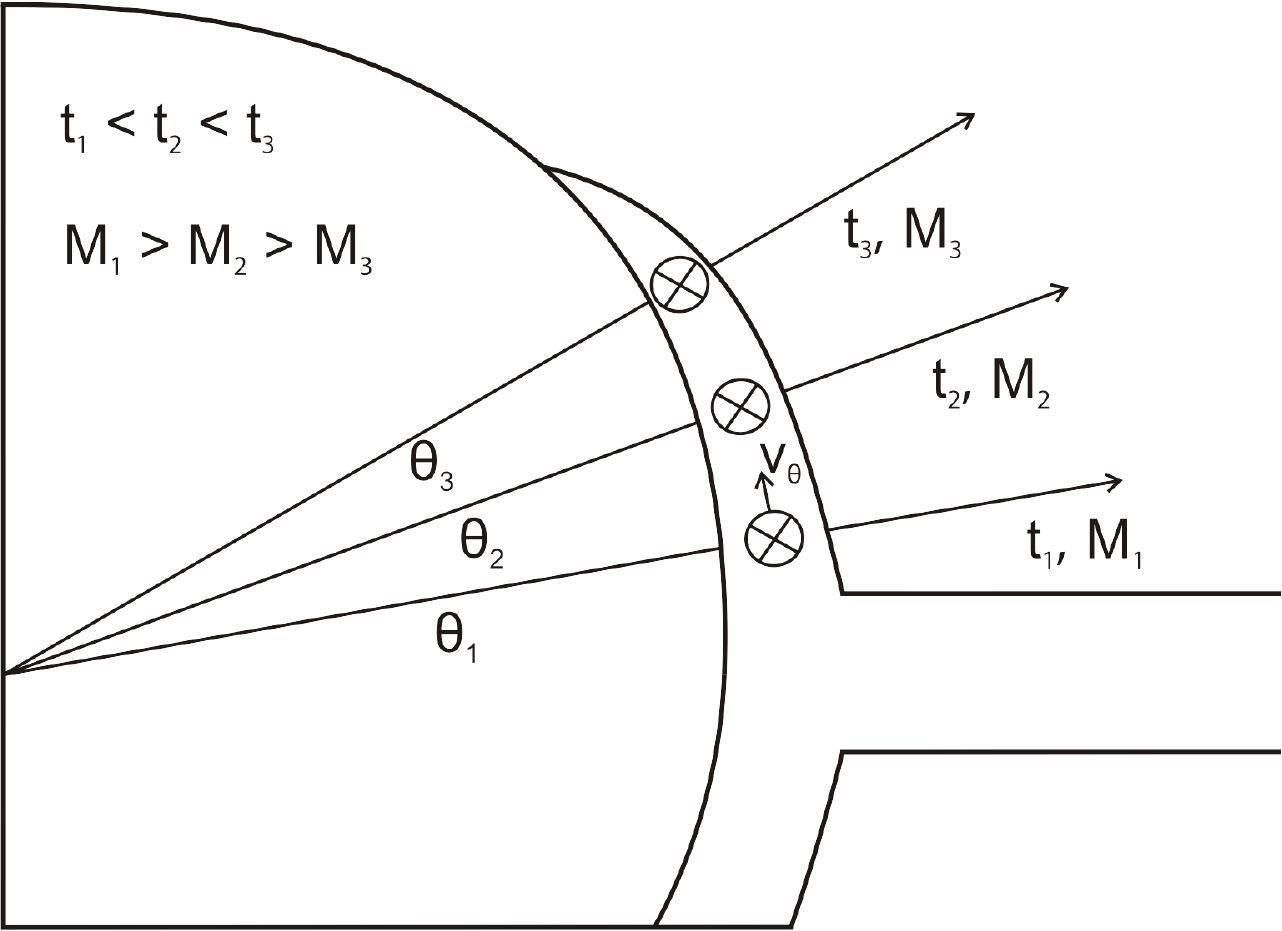}
\caption{Quasi-Lagrangian interpretation of the global SL evolution: as times goes by the simulation box drifts to higher latitudes and SL decelerates. See \S \ref{sect:global_outlook} for details.}
\label{fig:SL_lagrangian}
\end{figure}

On the other hand, this apparent SL evolution can be usefully interpreted in a {\it quasi-Lagrangian sense}, by effectively assuming that the simulation box {\it drifts to higher latitudes} as time goes by and SL decelerates. Figure \ref{fig:SL_lagrangian} shows schematic illustration of this correspondence, with longer time $t$ corresponding to higher $\theta$ and resulting in lower $M$. Of course, the precise rate at which this latitudinal drift would occur is impossible to predict without the explicit 3D calculation. Moreover, this drift could result in the variation of effective gravity (if the star is spinning), as well as thermodynamic conditions (e.g. cooling time), which are fixed in our setup. Nevertheless, we believe that this time-latitude ($t\to \theta$) analogy can provide at least a useful qualitative interpretation of the time evolution of the SL. Below we comment on its consequences for the case of accretion onto the WD.

This analogy predicts that the SL flow should stay supersonic at near-equatorial latitudes, steadily decelerating as the gas climbs towards the poles. Then one expects angular momentum and energy transport in the near-equatorial part of the SL to be dominated by the acoustic modes, like in the early stages of Figures \ref{fig2} \& \ref{fig110}. Mixing between the WD core and the SL should be minimal in this region, see \S \ref{sect:mix}. Angular momentum carried by the waves propagating into the deeper layers of the WD could be dissipated at significant depth (we refrain from identifying possible dissipation mechanisms in this work), spinning up these deep layers and heating them in a non-local fashion.

As the flow slows down during its meridional drift, at some latitude the SL would ultimately become transonic, quenching excitation of acoustic modes. The flow would then develop KH instability, giving rise to efficient turbulent mixing with the underlying layers of the WD. The surface of the WD would be directly spun-up and heated there by turbulent dissipation. This will happen predominantly at {\it high latitudes}, which may have important implications discussed in \S \ref{sect:bipolar}.

In global spherical geometry of the SL, we expect the propagating sonic wave pattern visible in our simulations to result in periodic modulation of the emission of the accreting object, see \citet{Belyaev1,Belyaev2}. The natural observational manifestations of this modulation could be the dwarf novae oscillations in the case of the WDs and quasi-periodic oscillations in the case of NSs, for which the SL origin has been previously proposed \citep{PiroB04,Molkov,Gilfanov,Revnivtsev}. Future global simulations that account for radiation transfer will be able to directly connect the flow behavior in the SL with observations of periodicity of compact objects.


\subsubsection{Bipolar nova outflows}
\label{sect:bipolar}

A growing body of observational evidence suggests that many nova ejectae are highly bipolar. This conclusion is supported by their optical/IR morphology \citep{Gill,Woudt,Chesneau,Chesneau12}, spectral line shape modeling \citep{Ribeiro}, polarization studies \citep{Kawabata}, and VLBI imaging \citep{Yang} of ejected nova material. The most common explanation for this asymmetry is the interaction of the ejected shell with the companion \citep{Livio,Lloyd,Porter}.

Here we speculate that such asymmetry may also naturally arise in the qualitative picture of the SL evolution advanced in \S \ref{sect:global_outlook}. Indeed, in this model most of the mixing between the SL and underlying WD material occurs at high latitudes, closer to the polar regions of the WD (with regard to the accretion disk plane). This should lead to more efficient injection of C \& O into the accreted gas further from the plane of the disk, leading to locally enhanced energy release per unit mass  during the thermonuclear runaway. As a result, ejection of material would be more energetic at some {\it higher} latitudes, naturally resulting in the bipolar morphology of the ejecta. 

The two testable predictions following from this idea of latitudinally dependent mixing are (1) the inhomogeneous (latitudinally-dependent) contamination of the ejected nova shells with dredged up C,N,O (weaker pollution near the equatorial region of the ejected shell) and (2) the alignment of the ejected shell with the binary angular momentum, perpendicular to which the disk equatorial plane is expected to be. The latter, however, is also expected in the standard picture of the ejecta interaction with the companion. Moreover, \citet{Porter} showed that rotation of the exploding layer on the WD surface significantly affects the asymmetry of the ejecta. Such (differential) rotation would naturally arise in the SL picture.

Needless to say, our suggestion for driving the ejecta asymmetry necessarily requires that SL does indeed form on the WD surface, which may require special circumstances such as the high enough mass accretion rate\footnote{Note, however, that determination of critical $\dot M$ in \citep{Piro} assumed that SL structure is mediated by {\it local} $\alpha$-viscosity.} $\dot M\gtrsim 10^{-8}$ M$_\odot$ yr$^{-1}$ \citep{Piro}. According to \citet{Fujimoto,Kahabka} nova explosions should still be possible at this $\dot M$.


\subsection{Comparison with previous studies}
\label{sect:compare}


It is important to put our study in proper context with regard to other work on boundary layers, SLs, transport and mixing. It should be considered as a natural extension of a series of studies by \citet{Belyaev1,Belyaev2,Belyaev3}, which were the first to identify sonic instability as the origin of the angular momentum transport in the BLs of accretion disks and provided its detailed characterization. Now we extend this approach to the SL morphology of accretion, while also studying mixing and effects of the non-trivial gas thermodynamics. Inclusion of the latter aspect makes our approach similar to \citet{Kley} who incorporated approximate radiation transport in their 2D simulations of the BLs. Our geometric setup closely parallels that of \citet{Belyaev0}, however, that study neglected vertical gravity and density stratification and used only isothermal EOS.

Acoustic waves have previously been claimed to emerge in axisymmetric models of the BLs computed in \citet{Godon} using $\alpha$-parametrization of the effective viscosity. Because of the azimuthal symmetry of these calculations such modes cannot drive the angular momentum transport and must be different from the intrinsically non-axisymmetric sonic waves seen in our simulations.

Unlike the theoretical work of \citet{InogamovSunyaev} and \citet{Piro} our study does not postulate some form of local effective $\alpha$-viscosity to operate in the SL. Instead, our goal was to {\it identify} the mechanism of the angular momentum transport in the Sl, and we demonstrate it to be accomplished by acoustic modes. Their non-local damping makes wave-mediated transport intrinsically non-local, precluding its description via some form of local effective viscosity.

There is also a body of literature devoted to understanding mixing of accreted material with underlying layers of the WD prior to novae or Type Ia SNe explosions, a subject we discussed in \S \ref{sect:mix}. These theoretical studies explored different instabilities that could result in local turbulence and mixing. The candidate mechanisms include convection \citep{Shankar,Glasner,Glasner14}, shear-driven KH instability \citep{Casanova10,Casanova11}, and breaking of gravity waves \citep{Rosner,Alexakis01,Alexakis04} driven by the resonant wind-wave instability \citep{Miles}. However, all these studies focused on the regime of subsonic shear and can thus be applicable only to the late stages of the SL evolution, when the flow is no longer supersonic. This is also the reason why these studies did not uncover the existence of the sonic instability and transport associated with it, which {\it are the key findings of} our work.


\section{Conclusions}
\label{sect:conclusions}


This work explores physics of the supersonic spreading layers of astrophysical accretion disks around compact objects such as the white dwarfs and neutron stars. It addresses the important issues of angular momentum and energy transport, mixing and thermodynamics in the relevant geometric setup.  

Our main findings can be summarized as follows.
\begin{itemize}

\item
We universally find excitation of the sonic instability in our simulations, which saturates and drives large-scale wave motions dominating the evolution of the system in the supersonic regime. This finding is insensitive to the assumed thermodynamics of the flow, boundary conditions, etc. and naturally extends the results of \citet{Belyaev1,Belyaev2,Belyaev3} to the SL morphology.

\item
{{In our simulations of the supersonic spreading layer the transport of momentum and energy by acoustic waves dominates over other forms of the transport, e.g. molecular or numerical viscosity}.} This mode of transport is inherently non-local in nature and cannot be characterized via some local effective viscosity (such as e.g. commonly adopted $\alpha$-viscosity prescription). Non-local transport of energy should have important implications for triggering of the Type I X-ray bursts in LMXBs and understanding the thermal state of accreting NSs. 

\item 
Thermodynamics of the flow determines the efficiency of the wave-driven transport. It also plays important role in evolution of the SL, causing significant rearrangement of its properties when cooling is not fast enough.

\item
In the supersonic regime mixing between the SL and underlying layers is not efficient since the momentum transport in the SL is driven by the wave-like motions. Mixing intensifies when the flow decelerates, becomes transonic and susceptible to KH instability. 

\item
In our 2D setup without explicit forcing the back-reaction of the modes on the flow (dissipation in weak shocks) results in SL deceleration. We provide a qualitative interpretation of this evolution in terms of the global meridional spreading of matter as the SL slows down.

\item 
Our global picture of the SL evolution predicts weak material mixing in the near-equatorial, supersonic regions, and strong mixing at higher latitudes, where the SL becomes transonic and KH-unstable. We speculate on the possible relevance of such latitudinally inhomogeneous mixing for producing bipolar morphology of nova ejecta.

\end{itemize}

These results should be tested with future global, 3D simulations of the SL structure and evolution incorporating realistic physics (e.g. radiation transfer, transition from the boundary layer in the disk to the SL, etc.). {{In particular, global models may reveal if any other mechanism
competitive with acoustic modes  in efficiency of angular momentum and energy transport could operate in the spreading layer.}}


\acknowledgments

We thank Anthony Piro for discussions and Mike Belyaev for useful comments on the manuscript. We are grateful to an anonymous referee for a number of useful criticisms and suggestions. RRR is an IBM Einstein Fellow at the IAS. He thanks Lebedev Physical Institute for hospitality during the final stages of this work, and acknowledges financial support by NSF via grant AST-1409524, NASA via grant 14-ATP14-0059, and The Ambrose Monell Foundation. Simulations presented in this article used computational resources supported by PICSciE-OIT TIGRESS High Performance Computing Center.


\bibliography{mybib}

\begin{thebibliography}{63}
\expandafter\ifx\csname natexlab\endcsname\relax\def\natexlab#1{#1}\fi

\bibitem[{{Alexakis} {et~al.}(2004){Alexakis}, {Calder}, {Heger}, {Brown},
  {Dursi}, {Truran}, {Rosner}, {Lamb}, {Timmes}, {Fryxell}, {Zingale},
  {Ricker}, \& {Olson}}]{Alexakis04}
{Alexakis}, A., {Calder}, A.~C., {Heger}, A., {Brown}, E.~F., {Dursi}, L.~J.,
  {Truran}, J.~W., {Rosner}, R., {Lamb}, D.~Q., {Timmes}, F.~X., {Fryxell}, B.,
  {Zingale}, M., {Ricker}, P.~M., \& {Olson}, K. 2004, \apj, 602, 931

\bibitem[{{Armitage}(2002)}]{Armitage}
{Armitage}, P.~J. 2002, \mnras, 330, 895

\bibitem[{{Balbus} \& {Hawley}(1991)}]{MRI}
{Balbus}, S.~A. \& {Hawley}, J.~F. 1991, \apj, 376, 214

\bibitem[{{Balsara} {et~al.}(2009){Balsara}, {Fisker}, {Godon}, \&
  {Sion}}]{Balsara09}
{Balsara}, D.~S., {Fisker}, J.~L., {Godon}, P., \& {Sion}, E.~M. 2009, \apj,
  702, 1536

\bibitem[{{Belyaev} \& {Rafikov}(2012)}]{Belyaev0}
{Belyaev}, M.~A. \& {Rafikov}, R.~R. 2012, \apj, 752, 115

\bibitem[{{Belyaev} {et~al.}(2012){Belyaev}, {Rafikov}, \& {Stone}}]{Belyaev1}
{Belyaev}, M.~A., {Rafikov}, R.~R., \& {Stone}, J.~M. 2012, \apj, 760, 22

\bibitem[{{Belyaev} {et~al.}(2013{\natexlab{a}}){Belyaev}, {Rafikov}, \&
  {Stone}}]{Belyaev2}
---. 2013{\natexlab{a}}, \apj, 770, 67

\bibitem[{{Belyaev} {et~al.}(2013{\natexlab{b}}){Belyaev}, {Rafikov}, \&
  {Stone}}]{Belyaev3}
---. 2013{\natexlab{b}}, \apj, 770, 68

\bibitem[{{Casanova} {et~al.}(2010){Casanova}, {Jos{\'e}},
  {Garc{\'{\i}}a-Berro}, {Calder}, \& {Shore}}]{Casanova10}
{Casanova}, J., {Jos{\'e}}, J., {Garc{\'{\i}}a-Berro}, E., {Calder}, A., \&
  {Shore}, S.~N. 2010, \aap, 513, L5

\bibitem[{{Casanova} {et~al.}(2011){Casanova}, {Jos{\'e}},
  {Garc{\'{\i}}a-Berro}, {Calder}, \& {Shore}}]{Casanova11}
---. 2011, \aap, 527, A5

\bibitem[{{Chandrasekhar}(1960)}]{Chandra}
{Chandrasekhar}, S. 1960, Proceedings of the National Academy of Science, 46,
  253

\bibitem[{{Chesneau} {et~al.}(2012){Chesneau}, {Lagadec}, {Otulakowska-Hypka},
  {Banerjee}, {Woodward}, {Harvey}, {Spang}, {Kervella}, {Millour}, {Nardetto},
  {Ashok}, {Barlow}, {Bode}, {Evans}, {Lynch}, {O'Brien}, {Rudy}, \&
  {Russell}}]{Chesneau12}
{Chesneau}, O., {Lagadec}, E., {Otulakowska-Hypka}, M., {Banerjee}, D.~P.~K.,
  {Woodward}, C.~E., {Harvey}, E., {Spang}, A., {Kervella}, P., {Millour}, F.,
  {Nardetto}, N., {Ashok}, N.~M., {Barlow}, M.~J., {Bode}, M., {Evans}, A.,
  {Lynch}, D.~K., {O'Brien}, T.~J., {Rudy}, R.~J., \& {Russell}, R.~W. 2012,
  \aap, 545, A63

\bibitem[{{Chesneau} {et~al.}(2011){Chesneau}, {Meilland}, {Banerjee}, {Le
  Bouquin}, {McAlister}, {Millour}, {Ridgway}, {Spang}, {ten Brummelaar},
  {Wittkowski}, {Ashok}, {Benisty}, {Berger}, {Boyajian}, {Farrington},
  {Goldfinger}, {Merand}, {Nardetto}, {Petrov}, {Rivinius}, {Schaefer},
  {Touhami}, \& {Zins}}]{Chesneau}
{Chesneau}, O., {Meilland}, A., {Banerjee}, D.~P.~K., {Le Bouquin}, J.-B.,
  {McAlister}, H., {Millour}, F., {Ridgway}, S.~T., {Spang}, A., {ten
  Brummelaar}, T., {Wittkowski}, M., {Ashok}, N.~M., {Benisty}, M., {Berger},
  J.-P., {Boyajian}, T., {Farrington}, C., {Goldfinger}, P.~J., {Merand}, A.,
  {Nardetto}, N., {Petrov}, R., {Rivinius}, T., {Schaefer}, G., {Touhami}, Y.,
  \& {Zins}, G. 2011, \aap, 534, L11

\bibitem[{{Deibel} {et~al.}(2015){Deibel}, {Cumming}, {Brown}, \&
  {Page}}]{MAXI}
{Deibel}, A., {Cumming}, A., {Brown}, E.~F., \& {Page}, D. 2015, ArXiv e-prints

\bibitem[{{Dong} {et~al.}(2011){Dong}, {Rafikov}, \& {Stone}}]{Dong}
{Dong}, R., {Rafikov}, R.~R., \& {Stone}, J.~M. 2011, \apj, 741, 57

\bibitem[{{Fisker} \& {Balsara}(2005)}]{Balsara05}
{Fisker}, J.~L. \& {Balsara}, D.~S. 2005, \apjl, 635, L69

\bibitem[{{Fujimoto}(1982)}]{Fujimoto}
{Fujimoto}, M.~Y. 1982, \apj, 257, 767

\bibitem[{{Gehrz} {et~al.}(1998){Gehrz}, {Truran}, {Williams}, \&
  {Starrfield}}]{Gehrz}
{Gehrz}, R.~D., {Truran}, J.~W., {Williams}, R.~E., \& {Starrfield}, S. 1998,
  \pasp, 110, 3

\bibitem[{{Ghosh} \& {Lamb}(1978)}]{Ghosh}
{Ghosh}, P. \& {Lamb}, F.~K. 1978, \apjl, 223, L83

\bibitem[{{Gilfanov} \& {Revnivtsev}(2005)}]{Gilfanov}
{Gilfanov}, M. \& {Revnivtsev}, M. 2005, Astronomische Nachrichten, 326, 812

\bibitem[{{Gilfanov} {et~al.}(2003){Gilfanov}, {Revnivtsev}, \&
  {Molkov}}]{Molkov}
{Gilfanov}, M., {Revnivtsev}, M., \& {Molkov}, S. 2003, \aap, 410, 217

\bibitem[{{Gill} \& {O'Brien}(2000)}]{Gill}
{Gill}, C.~D. \& {O'Brien}, T.~J. 2000, \mnras, 314, 175

\bibitem[{{Glasner} {et~al.}(2014){Glasner}, {Livne}, \& {Truran}}]{Glasner14}
{Glasner}, A.~S., {Livne}, E., \& {Truran}, J.~W. 2014, in Astronomical Society
  of the Pacific Conference Series, Vol. 490, Stell Novae: Past and Future
  Decades, ed. P.~A. {Woudt} \& V.~A.~R.~M. {Ribeiro}, 295

\bibitem[{{Glasner} \& {Livne}(1995)}]{Glasner}
{Glasner}, S.~A. \& {Livne}, E. 1995, \apjl, 445, L149

\bibitem[{{Glatzel}(1988)}]{Glatzel}
{Glatzel}, W. 1988, \mnras, 231, 795

\bibitem[{{Godon}(1995)}]{Godon}
{Godon}, P. 1995, \mnras, 274, 61

\bibitem[{{Hertfelder} \& {Kley}(2015)}]{Kley}
{Hertfelder}, M. \& {Kley}, W. 2015, ArXiv e-prints

\bibitem[{{Hertfelder} {et~al.}(2013){Hertfelder}, {Kley}, {Suleimanov}, \&
  {Werner}}]{Hertfelder13}
{Hertfelder}, M., {Kley}, W., {Suleimanov}, V., \& {Werner}, K. 2013, \aap,
  560, A56

\bibitem[{{Inogamov} \& {Sunyaev}(1999)}]{InogamovSunyaev}
{Inogamov}, N.~A. \& {Sunyaev}, R.~A. 1999, Astronomy Letters, 25, 269

\bibitem[{{Inogamov} \& {Sunyaev}(2010)}]{IS2}
---. 2010, Astronomy Letters, 36, 848

\bibitem[{{Kahabka} \& {van den Heuvel}(1997)}]{Kahabka}
{Kahabka}, P. \& {van den Heuvel}, E.~P.~J. 1997, \araa, 35, 69

\bibitem[{{Kawabata} {et~al.}(2006){Kawabata}, {Ohyama}, {Ebizuka}, {Takata},
  {Yoshida}, {Isogai}, {Norimoto}, {Okazaki}, \& {Saitou}}]{Kawabata}
{Kawabata}, K.~S., {Ohyama}, Y., {Ebizuka}, N., {Takata}, T., {Yoshida}, M.,
  {Isogai}, M., {Norimoto}, Y., {Okazaki}, A., \& {Saitou}, M.~S. 2006, \aj,
  132, 433

\bibitem[{{Kley}(1989)}]{Kley89}
{Kley}, W. 1989, \aap, 222, 141

\bibitem[{{Livio} {et~al.}(1990){Livio}, {Shankar}, {Burkert}, \&
  {Truran}}]{Livio}
{Livio}, M., {Shankar}, A., {Burkert}, A., \& {Truran}, J.~W. 1990, \apj, 356,
  250

\bibitem[{{Lloyd} {et~al.}(1997){Lloyd}, {O'Brien}, \& {Bode}}]{Lloyd}
{Lloyd}, H.~M., {O'Brien}, T.~J., \& {Bode}, M.~F. 1997, \mnras, 284, 137

\bibitem[{{Lynden-Bell} \& {Pringle}(1974)}]{LyndenBell74}
{Lynden-Bell}, D. \& {Pringle}, J.~E. 1974, \mnras, 168, 603

\bibitem[{{Miles}(1957)}]{Miles}
{Miles}, J.~W. 1957, Journal of Fluid Mechanics, 3, 185

\bibitem[{{Miles}(1958)}]{Miles1958}
---. 1958, Journal of Fluid Mechanics, 4, 538

\bibitem[{{Narayan} \& {Popham}(1993)}]{NarPop}
{Narayan}, R. \& {Popham}, R. 1993, \nat, 362, 820

\bibitem[{{Piro}(2015)}]{PiroWDmix}
{Piro}, A.~L. 2015, \apj, 801, 137

\bibitem[{{Piro} \& {Bildsten}(2004{\natexlab{a}})}]{PiroB04}
{Piro}, A.~L. \& {Bildsten}, L. 2004{\natexlab{a}}, \apjl, 616, L155

\bibitem[{{Piro} \& {Bildsten}(2004{\natexlab{b}})}]{Piro}
---. 2004{\natexlab{b}}, \apj, 610, 977

\bibitem[{{Popham} \& {Narayan}(1995)}]{PophamWD}
{Popham}, R. \& {Narayan}, R. 1995, \apj, 442, 337

\bibitem[{{Popham} {et~al.}(1993){Popham}, {Narayan}, {Hartmann}, \&
  {Kenyon}}]{PophamPMS}
{Popham}, R., {Narayan}, R., {Hartmann}, L., \& {Kenyon}, S. 1993, \apjl, 415,
  L127

\bibitem[{{Porter} {et~al.}(1998){Porter}, {O'Brien}, \& {Bode}}]{Porter}
{Porter}, J.~M., {O'Brien}, T.~J., \& {Bode}, M.~F. 1998, \mnras, 296, 943

\bibitem[{{Pringle}(1977)}]{Pringle77}
{Pringle}, J.~E. 1977, \mnras, 178, 195

\bibitem[{{Regev}(1983)}]{Regev83}
{Regev}, O. 1983, \aap, 126, 146

\bibitem[{{Revnivtsev} \& {Gilfanov}(2006)}]{Revnivtsev}
{Revnivtsev}, M.~G. \& {Gilfanov}, M.~R. 2006, \aap, 453, 253

\bibitem[{{Ribeiro} {et~al.}(2013){Ribeiro}, {Munari}, \& {Valisa}}]{Ribeiro}
{Ribeiro}, V.~A.~R.~M., {Munari}, U., \& {Valisa}, P. 2013, \apj, 768, 49

\bibitem[{{Rosner} {et~al.}(2001{\natexlab{a}}){Rosner}, {Alexakis}, {Young},
  {Truran}, \& {Hillebrandt}}]{Alexakis01}
{Rosner}, R., {Alexakis}, A., {Young}, Y.-N., {Truran}, J.~W., \&
  {Hillebrandt}, W. 2001{\natexlab{a}}, \apjl, 562, L177

\bibitem[{{Rosner} {et~al.}(2001{\natexlab{b}}){Rosner}, {Alexakis}, {Young},
  {Truran}, \& {Hillebrandt}}]{Rosner}
---. 2001{\natexlab{b}}, \apjl, 562, L177

\bibitem[{{Shankar} \& {Arnett}(1994)}]{Shankar}
{Shankar}, A. \& {Arnett}, D. 1994, \apj, 433, 216

\bibitem[{{Starrfield}(2015)}]{sn1a}
{Starrfield}, S. 2015, ArXiv e-prints

\bibitem[{{Steinacker} \& {Papaloizou}(2002)}]{Steinacker}
{Steinacker}, A. \& {Papaloizou}, J.~C.~B. 2002, \apj, 571, 413

\bibitem[{{Stone} {et~al.}(2008){Stone}, {Gardiner}, {Teuben}, {Hawley}, \&
  {Simon}}]{StoneAthena}
{Stone}, J.~M., {Gardiner}, T.~A., {Teuben}, P., {Hawley}, J.~F., \& {Simon},
  J.~B. 2008, \apjs, 178, 137

\bibitem[{{Strohmayer} \& {Bildsten}(2006)}]{Stro}
{Strohmayer}, T. \& {Bildsten}, L. {New views of thermonuclear bursts}, ed.
  W.~H.~G. {Lewin} \& M.~{van der Klis}, 113--156

\bibitem[{{Suleimanov} {et~al.}(2014){Suleimanov}, {Hertfelder}, {Werner}, \&
  {Kley}}]{Suleimanov14}
{Suleimanov}, V., {Hertfelder}, M., {Werner}, K., \& {Kley}, W. 2014, \aap,
  571, A55

\bibitem[{{Truran}(1982)}]{Truran82}
{Truran}, J.~W. 1982, in Essays in Nuclear Astrophysics, ed. C.~A. {Barnes},
  D.~D. {Clayton}, \& D.~N. {Schramm}, 467

\bibitem[{{Truran} \& {Livio}(1986)}]{Truran}
{Truran}, J.~W. \& {Livio}, M. 1986, \apj, 308, 721

\bibitem[{{Tylenda}(1981)}]{Tylenda81}
{Tylenda}, R. 1981, Acta Astronomica, 31, 267

\bibitem[{{Velikhov}(1959)}]{Velikhov}
{Velikhov}, E.~P. 1959, Soviet Physics JETP, 36, 1398

\bibitem[{{Woudt} {et~al.}(2009){Woudt}, {Steeghs}, {Karovska}, {Warner},
  {Groot}, {Nelemans}, {Roelofs}, {Marsh}, {Nagayama}, {Smits}, \&
  {O'Brien}}]{Woudt}
{Woudt}, P.~A., {Steeghs}, D., {Karovska}, M., {Warner}, B., {Groot}, P.~J.,
  {Nelemans}, G., {Roelofs}, G.~H.~A., {Marsh}, T.~R., {Nagayama}, T., {Smits},
  D.~P., \& {O'Brien}, T. 2009, \apj, 706, 738

\bibitem[{{Yang} {et~al.}(2015){Yang}, {Paragi}, {O'Brien}, {Chomiuk}, \&
  {Linford}}]{Yang}
{Yang}, J., {Paragi}, Z., {O'Brien}, T.~J., {Chomiuk}, L., \& {Linford}, J.~D.
  2015, ArXiv e-prints

\end{thebibliography}

\end{document}